\documentclass[fleqn,usenatbib]{mnras}

\usepackage{newtxtext,newtxmath}

\usepackage[T1]{fontenc}
\usepackage{ae,aecompl}
\usepackage{graphicx}	
\usepackage{amsmath}	
\usepackage{amssymb}	

\usepackage{subcaption}  
\usepackage{url}

\usepackage [autostyle, english = british]{csquotes}  

\title[Automatic Tidal Feature Detection]{Identification of Low Surface Brightness Tidal Features in Galaxies Using Convolutional Neural Networks}

\author[M. Walmsley et al.]{
Mike Walmsley,$^{1, 2}$\thanks{E-mail: mike.walmsley@physics.ox.ac.uk (MW)}
Annette M. N. Ferguson$^{3}$
Robert G. Mann,$^{3}$
Chris. J. Lintott$^{2}$
\\  
$^{1}$School of Physics and Astronomy, James Clark Maxwell Building, Peter Guthrie Tait Road, Edinburgh EH9 3FD, United Kingdom \\
$^{2}$Oxford Astrophysics, The Denys Wilkinson Building, Oxford OX1 3RH, United Kingdom \\
$^{3}$Institute for Astronomy, Royal Observatory Edinburgh, University of Edinburgh, Blackford Hill, Edinburgh EH9 3HJ, United Kingdom
}

\date{Accepted XXX. Received YYY; in original form ZZZ}

\pubyear{2017}

\begin{document}
\label{firstpage}
\pagerange{\pageref{firstpage}--\pageref{lastpage}}
\maketitle

\begin{abstract}
Faint tidal features around galaxies record their merger and interaction histories over cosmic time. Due to their low surface brightnesses and complex morphologies, existing automated methods struggle to detect such features and most work to date has heavily relied on visual inspection. This presents a major obstacle to quantitative study of tidal debris features in large statistical samples,  and hence the ability to be able to use these features to advance understanding of the galaxy population as a whole.  This paper uses convolutional neural networks (CNNs) with dropout and augmentation to identify galaxies in the CFHTLS-Wide Survey that have faint tidal features. Evaluating the performance of the CNNs against previously-published expert visual classifications,  we find that our method achieves high (76\%) completeness and low (20\%) contamination, and also performs considerably better than other automated methods recently applied in the literature.  We argue that CNNs offer a promising approach to effective automatic identification of low surface brightness tidal debris features in and around galaxies. When applied to  forthcoming deep wide-field imaging surveys (e.g.  LSST, Euclid), CNNs have the potential to provide a several order-of-magnitude increase in the sample size of morphologically-perturbed galaxies and thereby facilitate a much-anticipated revolution in terms of quantitative low surface brightness science. 
\end{abstract}




\section{Introduction}
\label{introduction}

Hierarchical models of galaxy formation suggest that present-day galaxies assemble their mass
through the repeated aggregation of smaller systems and through the smooth accretion of gas which fuels {\it in situ} star formation (e.g. \citealt{White1991, Abadi2002}).  While it is generally agreed that the most massive galaxies have acquired almost all of their stars through mergers, the relative contribution of {\it in situ} star formation and directly accreted stellar mass remains an open question across much of the galaxy mass spectrum (e.g. \citealt{Rodriguez-Gomez2016, Qu2017, Lee2017, Fitts2018}). Furthermore, the rates of major (mass ratio $\geq$1:4) and minor merger events, and their role in shaping various galaxy components, are also not yet well understood (e.g. \citealt{Lotz2011,Martin2017,Lofthouse2017}). 

Given the intrinsic uncertainties in {\it ab initio} modelling galaxy formation within a cosmological context (see the discussion in \citealt{Hopkins2018}), a purely empirical measure of the frequency and nature of galaxy mergers and accretions is highly desirable.  It is well established that such events leave long-lasting observational signatures in the form of low surface brightness tidal streams, shells and perturbations (e.g. \citealt{Toomre1972, Quinn1984, Cooper2010}). In galaxy outskirts, where the dynamical timescales are several gigayears or longer, these features are predicted to be particularly apparent \citep{Johnston1996, Cooper2013}. Indeed, much stellar substructure of this nature has already been detected in the peripheral regions of the Milky Way (e.g. \citealt{Belokurov2006}), M31 (see the review by \cite{Ferguson2016}), and other nearby galaxies (e.g. \citealt{Martinez-Delgado2010,Duc2015}). Low surface brightness tidal features are therefore a powerful means to identify systems which have undergone recent mergers and accretions. The morphology and properties of these features hold vital clues to the nature of the events which have created them \citep{Hendel2015,Pop2018}. 

One of the main obstacles in such studies is the difficulty in reliably identifying faint tidal features. Part of this problem stems from the fact that morphological merger signatures only persist for a finite duration after an interaction has taken place, with the exact timescale dependent on the details of the orbital interaction as well as the properties of the host galaxy \citep{Lotz2008a}. Although predicted to be very common, minor mergers are particularly challenging to investigate because they generate faint signatures which are detectable over shorter timescales \citep{Lotz2011}. Indirect evidence for minor mergers from resulting morphological transformations (e.g. bulge growth) can provide sensitivity to events that have occurred over longer timescales but it is often difficult to distinguish these transformations from secular processes (e.g. \citealt{Hopkins2009,Kormendy2004,Conselice2003}). 

Another major challenge comes from the process of actually identifying the tidal features on deep galaxy images. Most work to date has focused on visual inspection of individual galaxies or relatively small samples (e.g. \citealt{Malin1983,Martinez-Delgado2010,Sheen2012}), often on images that have been specifically manipulated in order to enhance the appearance of low surface brightness features (e.g. \citealt{Miskolczi2011,Kado-Fong2018, Morales2018, Hood2018}).  However, many important questions about the role of interactions and mergers in driving galaxy evolution require large statistical samples (i.e. several thousand systems or more) for which expert human classification becomes impractical.  Unfortunately, there has been relatively little effort to date in devising automatic methods to detect and characterise low surface brightness emission in galaxies and the methods invoked are not particularly well-suited to detecting the faint tidal features typical of minor mergers.
Techniques may be broadly grouped into two categories -- those which rely on model subtraction and those which appeal to non-parametric feature extraction. 

Model subtraction methods work by removing the expected flux using a parametric light profile and then quantifying the 
amount of residual light (e.g. \citealt{VanDokkum2005,Tal2009,Adams2012}). This approach works best on galaxies with smooth radially-symmetric morphologies because of the difficulty in constructing light profiles that accurately model the basic morphology. Non-parametric feature extraction methods measure one or several hand-crafted image parameters thought to correlate with post-merger disruption (e.g. \citealt{Conselice2003,Lotz2004,Freeman2013,Pawlik2016}), and then apply selection cuts or machine learning estimation to identify the most likely candidates. These methods allow for a broader range of morphologies to be classified but can be easily confused by complex features such as spiral arms  (e.g. \citealt{Kartaltepe2010}) and are typically only sensitive to certain major merger stages \citep{Lotz2008, Lotz2011, Snyder2015}. 

Motivated by the desire to develop a more generalised method of tidal debris detection and classification, and one which can be applied to the specific problem of identifying (and ultimately characterising) faint features around galaxies in large statistical samples, we explore a new approach based on convolutional neural networks (hereafter CNNs).  Various authors (e.g. \citealt{Dieleman2015, Sanchez2017}) have demonstrated that CNNs can accurately classify general galaxy morphology and recently \cite{Ackermann2018} showed that CNNs can be used to identify merging galaxy pairs in SDSS Data Release 7 images \citep{Darg2010}.  Here, we use CNNs with dropout and augmentation to identify galaxies in the CFHTLS-Wide Survey that have faint tidal features in their outer regions. Through application to a galaxy sample that has been previously visually-searched for debris features, we demonstrate the reliability and effectiveness of our automated technique. We also show that its performance compares favourably to two other methods that have been recently applied in the literature.  

The paper is organized as follows.  In Section \ref{data} we describe the sample of galaxies under study. In Section \ref{single}, we describe the motivation for our approach and the design, training and performance of a single network, while in Section \ref{ensemble} we discuss the improvements obtained through using an ensemble of several networks.  In Section \ref{comparison_with_current_methods} we compare with the current approaches of WND-CHARM \citep{Shamir2012} and shape asymmetry \citep{Pawlik2016} and in Sections \ref{discussion} and \ref{conclusion} we discuss our results and conclusions. 

\section{Data}
\label{data}

We base our analysis on data products from the Wide component of the Canada-France-Hawaii Telescope Legacy Survey, hereafter CFHTLS-Wide \citep{Gwyn2012}.  This survey covers approximately 170 deg$^2$ of sky in four patches and uses filters \textit{u*}, \textit{g'}, \textit{r'}, \textit{i'}  and \textit{z'} with an exposure time of approximately one hour per filter per field. \cite{Atkinson2013} (hereafter A13) used visual classfications to study the incidence of faint tidal features in a sample of $\sim1800$ luminous galaxies drawn from this survey, making it an ideal sample against which to benchmark the performance of CNNs. 

The A13 sample contains 1781 galaxies that were selected to lie within the redshift range $0.04 < z < 0.2$ and to have magnitude $15.5 < r' < 17$. These cuts were adopted so as to allow for comparison with previous work on tidal feature classification, to minimise contamination from stars misidentified as galaxies and to limit the sample size to a manageable number for visual inspection.  As discussed in A13, this sample is heavily biased towards bright systems, with most galaxies lying the range $-23 < M_{r'}<-20$ mag.  The typical half-light radii of the galaxies is 2-6 arcsec. 

\begin{figure}
  \includegraphics[width=\columnwidth]{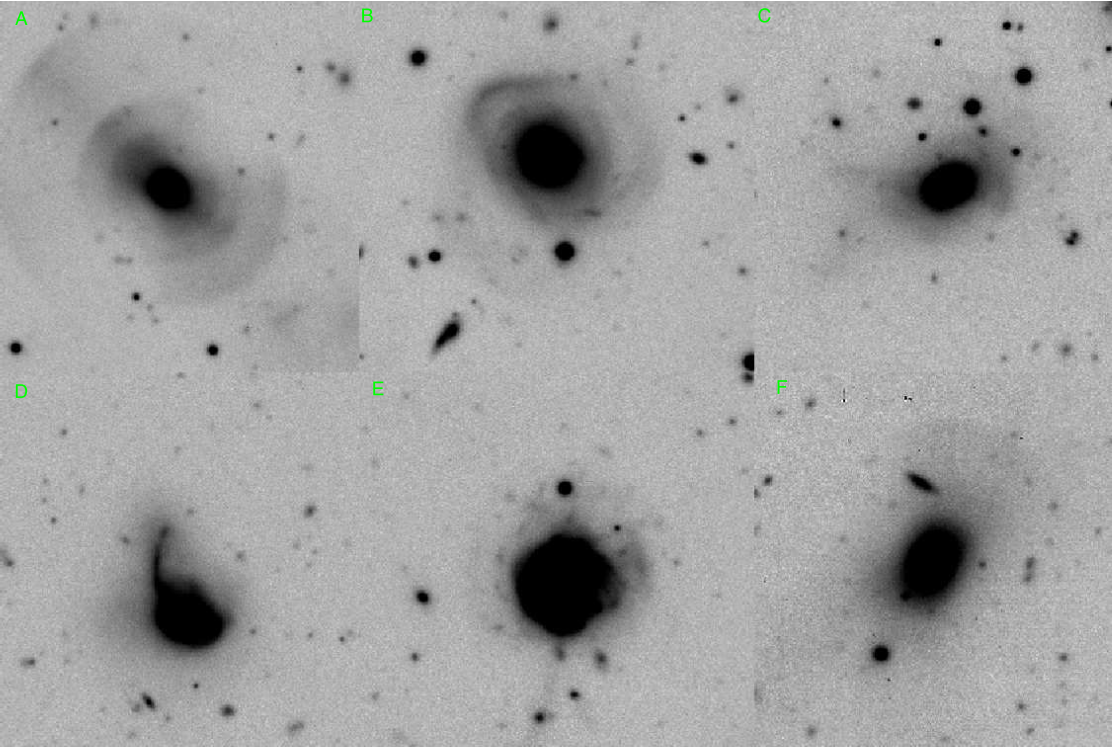}
  \caption{Tidal feature classes defined by Atkinson et al (2013). Clockwise from top left: shell (A), stream (B), misc. diffuse (C), arm (D), linear (E), fan (F). Reproduced by permission of the original authors and the AAS.}
  \label{atkinson_examples}
\end{figure}

The A13 study used thumbnails in the \textit{g'}, \textit{r'} and \textit{i'} bands as these were the highest signal-to-noise images. These thumbnails were 
stacked together to increase contrast. A13 estimate a limiting {\it g}-band surface brightness of $\approx 27.7$ mag arcsec$^{-2}$ over small scales. Each stacked image was visually inspected and placed into one of five categories depending on the confidence of the inspector that a tidal feature was present. These ranged from very high confidence of the presence of a feature (level four) to a feature with around 75\% certainty (level three) and so on, until very high confidence was reached that no tidal features were present to the depth of the data (level zero). If tidal structure was deemed to be present then it was further classified into six non-exclusive tidal feature classes -- shells, streams, miscellaneous diffuse structure, arms, linear features and broad fans.   Roughly 10\% of the A13 sample was classified independently by three experts to ensure that the visual classification scheme leads to consistent answers by multiple experts and is therefore reproducible. Following this, the entire sample was classified by a single inspector (Atkinson) in order to maximise consistency.  We consider these single expert labels as a ground truth against which to measure automated methods, and address reproducibility in Section \ref{discussion}. The archetypal examples provided by A13 of these feature classes are reproduced in Figure \ref{atkinson_examples}.

As the thumbnails utilised in the A13 study were not available to us, we had to recreate these from scratch, in an identical manner, so as to guarantee that our automated classifier had access to the same information as the human experts.   To this end, we extracted $256\times256$ pixel regions in the \textit{g'}, \textit{r'} and \textit{i'} bands around the galaxy centroid coordinates provided in the A13 catalogue using the CFHTLS cut-out service \citep{Erben2013}. These images were subsequently manipulated in a variety of ways, as will be described in Section \ref{preprocessing}.   

To reduce the complexity of the classification problem, tidal confidence labels were binned into binary classes. The choice to restrict the problem to a binary classification was motivated by the limited training data available (see Section \ref{network_architecture}) rather than any fundamental constraint. Non-tidal (0) was matched to confidence $\leq 25 \%$ (levels zero and one in the A13 scheme) whereas tidal (1) was matched to confidence  $\geq 75 \%$ (levels three and four in the A13 scheme). Galaxies with a tidal confidence of $50 \%$ were deemed to provide no useful information for our purpose and were cut from the sample. Of the 1781 galaxies in the original A13 sample, 24 could not be downloaded in all three bands from the CFHTLS cut-out service, giving an initial data sample of 1757 imaged galaxies. Of those, 1316 galaxies are re-labelled False (non-tidal) and 305 are re-labelled True (tidal). 136 have a confidence of 50\% and are therefore removed, leaving a final sample of 1621 galaxies with binary labels. The ability for the method to adapt to more subtle classes given sufficient training data is discussed in Section \ref{discussion}.

\section{Single Convolutional Neural Network Classifier}
\label{single}

\subsection{Introduction to Convolutional Neural Networks}
\label{introduction_to_networks}

CNNs are a subset of machine learning algorithms frequently used to identify patterns in tensors (i.e. n-dimensional arrays) where the spatial arrangement of values is important. Most commonly, these tensors are the pixel values of images. They routinely show state-of-the-art performance on various image classification benchmarks that require making discerning distinctions between classes and ignoring background effects \citep{Russakovsky2015}. We provide here a brief overview of how these methods work and refer the interested reader to \cite{LeCun2015} and references therein for detailed descriptions of CNNs, and to \cite{Dieleman2015,Lanusse2017} and \cite{Kim2017} for particular astrophysical applications. 

Neural networks are composed of repeated tensor operations called layers. The output of layer $l$, $\mathbf{x}^{l}$, is the input to layer $l+1$ and the arrangement and connectivity of layers is called the architecture. The net effect of a neural network is a non-linear mapping from input tensor to final layer output (i.e. prediction), with the aim being to learn the mapping which gives the true predictions.

Each type of layer performs a different operation. For example, the most basic is the fully-connected layer which performs the operation:

\begin{equation}
\label{neuralnetworklayer}
\mathbf{x}^{(l)} = f(\mathbf{w}^{(l)}\mathbf{x}^{(l-1)} + \mathbf{b}^{(l)})
\end{equation}

Consider the classification of an image using fully-connected layers. The image is encoded by the tensor of pixel values $\mathbf{x^{(0)}}$. This input propagates forward through the layers and is modified by the weights $\mathbf{w}^{(l)}$ and biases $\mathbf{b}^{(l)}$ of each layer through repeated operations of eqn. \ref{neuralnetworklayer}. The output of the final layer is interpreted as predictions for that image. 

\begin{figure}
  \includegraphics[width=\columnwidth]{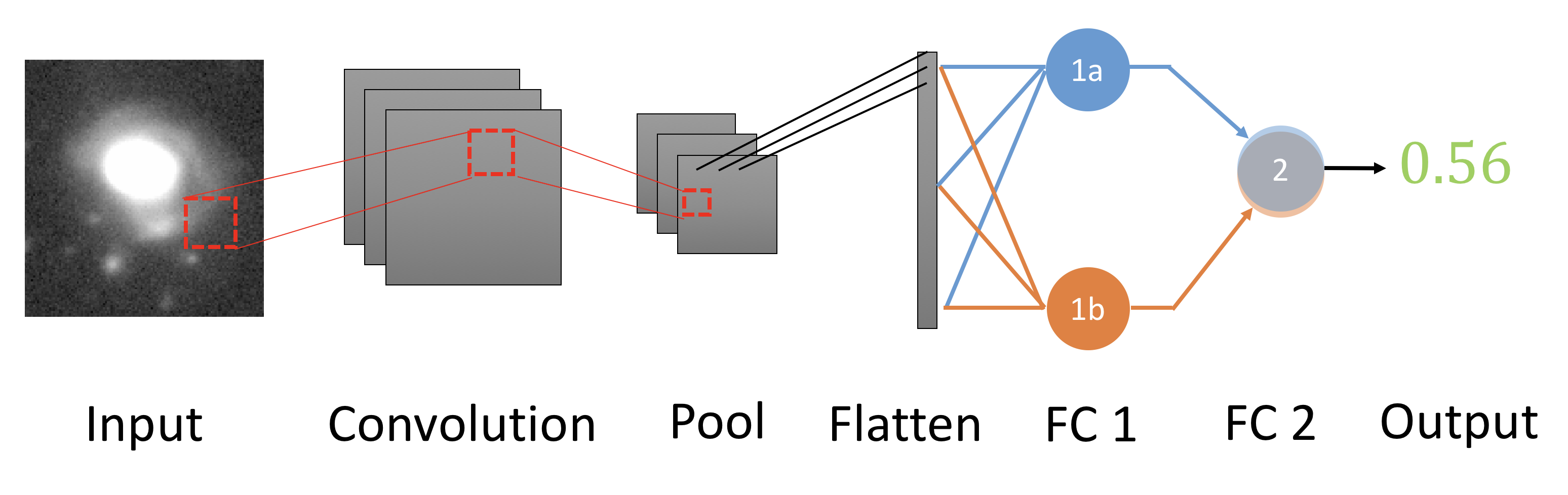}
  \caption{Illustrative diagram of a toy CNN. The pixel values of a galaxy image are taken as input. These are convolved with three filter matrices to create three feature maps. The feature maps are reduced in size by a pooling operation that preserves only the local 2x2 maxima, then 'flattened' and concatenated into one dimension. This flattened list of abstract features is the input for two fully-connected layers with two and one neurons respectively. The final fully-connected layer outputs a scalar value, to be interpreted as a 56\% confidence prediction of the galaxy having tidal features. In practice, the convolutional and pooling operations would repeat several times and the first fully-connected layer would include of order $100+$ neurons.}
  \label{cnn_illustration}
\end{figure}

Networks typically include two additional types of layer: convolutional and pooling. The convolutional layer operation can be described as 

\begin{equation}
\mathbf{x}_{n}^{(l)}=  f(\sum_{i}\mathbf{w}_{ij}^{(l)} \ast \mathbf{x}_{i}^{(l-1)} + \mathbf{b}_{i}^{(l)})
\end{equation}

\noindent where $\mathbf{w}_{ij}^{(l)}$ the filter of layer $l$. Convolutional layers identify features with a fixed scale relative to the filter size. On the other hand, pooling layers reduce the size of a feature map by aggregation, for example by preserving only the local 2x2 maxima (as in this work). When alternated with convolutional layers, pooling layers allow for features of increasing spatial scale to be detected.  Together, convolutional and pooling layers create increasingly abstract feature maps that encapsulate the image content. These features may then be classified using fully-connected layers. A toy CNN illustrating each operation is shown in Figure \ref{cnn_illustration}.  

The discriminative features to measure from the pixel data, and then to classify, are identified as part of the learning process, described in Section \ref{training}. This is in contrast to some other machine learning algorithms, such as random forests, which classify using user-defined image features like brightness and asymmetry.

We implement our network using the deep learning library Keras \citep{chollet2015keras}, with TensorFlow \citep{tensorflow2015-whitepaper} as a backend.

\subsection{Preprocessing}
\label{preprocessing}

For each galaxy in the sample, the thumbnails in each band were combined and manipulated in a variety of ways before being passed to the classifier. 
The "preprocessing" options investigated are listed below in order of operation.

\begin{enumerate}
\item \textbf{Aggregation} The \textit{g'}, \textit{r'} and \textit{i'} band images provide three tensors of pixel flux values, each of shape (height, width). These are combined to create a single tensor, which includes all pixel information on each galaxy, to be used as input to the network. The bands can be pixel-averaged to create a tensor of shape (height, width). Alternatively, the bands can be concatenated (i.e. placed next to one another) along a third colour dimension to create a tensor of shape (height, width, 3) in analogy with RGB images.

\item \textbf{Background estimation} This estimate is required for the pixel intensity clipping and masking procedures described below. To estimate the sky background, we use the functions \texttt{sigma\_clipped\_stats} and \texttt{make\_source\_mask} from the Python package Photutils (\citealt{Bradley2018}).

\texttt{sigma\_clipped\_stats} estimates background from the statistics of all unmasked pixels within a given $\sigma$ of the median unmasked pixel value. \texttt{sigma\_clipped\_stats} is called by \texttt{make\_source\_mask} to make an initial background estimate. \texttt{make\_source\_mask} then uses this estimate to detect and mask sources. The masked image is passed back to to \texttt{sigma\_clipped\_stats} for an updated background estimate. This procedure iterates five times, giving a final background estimate.

\item \textbf{Pixel intensity clipping}  The extreme intensity variation between the inner galaxy core and the tidal features can interfere with rescaling algorithms (see below). Retaining only pixels with intensities lower than 6$\sigma$ above the background avoids this issue. 

\item \textbf{Pixel intensity rescaling} Rescaling the pixels to reduce the dynamic range of the image ensures that the tidal features contribute to the first layer values. We apply to each tensor $x$ a rescaling mapping, for example $\sinh(x)$, $x^{a}$, or $\ln(x)$. Since the values of the first network layer are proportional to the input image pixel values, this avoids the untrained network initially seeing only the bright galaxy cores. 

\item \textbf{Masking}  The thumbnails have foreground and background objects, as well as occasional image artefacts, within the field-of-view. This introduces additional noise that could be mistaken for tidal features by the classifier. To mitigate this, pixels outside the contiguous galaxy light distribution can be masked. To identify which pixels to to mask, we use a combination of background estimation and mean convolutions to estimate which pixels are plausibly part of the galaxy. This process is described in detail in Section \ref{pawlik}.

\item \textbf{Local smoothing} This can enhance the appearance of faint tidal features near the signal-to-noise limit, albeit at the cost of a reduction in spatial resolution. We opt to replace each pixel with the local 3x3 average.
\end{enumerate}

As the optimal combination of these various preprocessing options for tidal feature detection is not initially obvious, we approach this problem empirically, by using a grid search - see Section \ref{gridsearch}.

\subsection{Training and Evaluation}
\label{training}

\begin{figure}
  \includegraphics[width=\columnwidth]{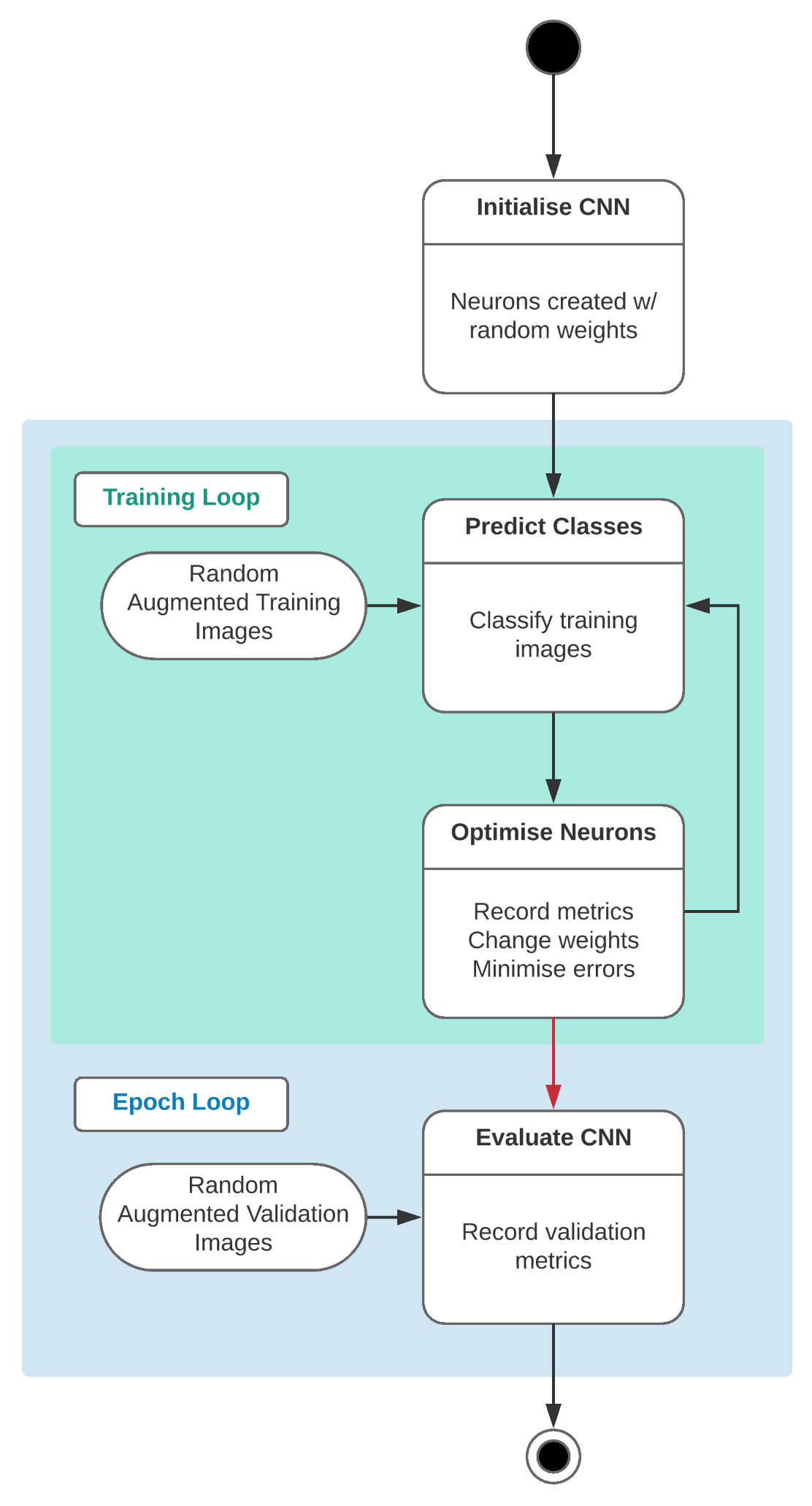}
  \caption{Flow chart of each stage for a single CNN. Red arrows denote steps which only occur after specified iterations have elapsed.}
  \label{cnn_flow}
\end{figure}

The CNN algorithm consists of two nested loops: an inner training loop and an outer epoch loop. The complete algorithm is illustrated in Figure \ref{cnn_flow}. 

With every iteration of the training loop, the network is gradually fit to the training data. 
A batch of unique labelled images (see Section \ref{augmentation}) is given as input to the CNN, and the CNN returns predictions.
The quality of these predictions is measured using a loss function.
For binary classification problems, a standard choice is the binary cross-entropy

\begin{equation}
  \label{loss}
  \mathcal{L} = - \sum^N_{i=1} y_i\log{p_i} + (1-y_i)\log(1-p_i)
\end{equation}

where the loss is summed over a batch of images of size $N$, each with a true label $y_i$ and a model score $p_i$.

The gradient of the loss function with respect to the weights and biases is computed, and the weights and biases are then updated to minimise the loss function.
The loop then repeats for a new batch of labelled images. 
Once a specified number of training loops have elapsed, the epoch loop is executed.

For every iteration of the epoch loop, the  network makes predictions for a batch of `unseen' validation images not used in the training process. Metrics for the quality of these predictions (for example, the mean accuracy) are recorded. The training process (i.e. multiple training loops) is then restarted with a new configuration. Once a specified number of epoch loops elapse, the algorithm terminates. 

All figures in this work use a batch size of 75 images, identified as the optimal number by the grid search (see Section \ref{gridsearch}). One epoch is arbitrarily set as 14 batches or 1050 training images. Batch images are randomly selected without replacement (i.e. selected only once) in equal proportion from the tidal and non-tidal galaxy training subsets. Once all images from a subset have been selected once and removed, the subset is refilled. Firstly, this approach provides the network with sufficient tidal examples to learn from. Secondly, it allows the network prediction to be interpreted as the probability that a given image is tidal and not merely a reflection of the base rate (i.e. the relative number of tidal versus non-tidal galaxy training examples seen by the network). Excluding the base rate during training ensures that predictions on a new sample will not be biased towards the training base rate. Each selected image is randomly transformed to augment the dataset (see Section \ref{augmentation}).

Any initial partitioning of data into training and validation images is arbitrary; we could have selected any set of images as validation images. We therefore need to check if the network is fortuitously performing better on those validation images than it would on a large set of new data. Smaller datasets are particularly susceptible to such accidental overperformance, as small number statistics make this scenario more likely.
We use five-fold cross validation to ensure our prediction quality metrics do not depend on an arbitrary division of data into training and validation subsets. In $n$-fold cross-validation, the complete data sample is split into $n$ random subsets. $n-1$ are used to train the classifier from scratch, and the remaining subset is used as validation data. This is repeated for all $n$ permutations. All prediction quality metrics in this work are then averaged from each of the five-fold cross-validation runs.

\subsection{Network Architecture}
\label{network_architecture}

A significant challenge with our CNN approach to tidal feature identification is our exceptionally small training sample.
Because every neuron connection is assigned a weight, CNNs typically have $> 10^{5}$ free parameters to learn (i.e. to fit to the data). Having many free parameters allows the learning of more complex features, but increases the entropic capacity of the classifier. Without a correspondingly large training sample to provide constraints, overfitting occurs and degrades the performance. CNNs are typically applied to samples of $10^{4}$ to $10^{10}$ images - see for example, \cite{Simonyan2014, Dieleman2015,Huertas-Company2015a,Kim2017} and \cite{Petrillo2017} - while our CFHTLS-Wide sample contains only 1621 galaxies, of which a mere 305 have tidal features. We therefore need to operate approximately two orders-of-magnitude below the minimum sample sizes normally used by CNNs.  

\begin{figure}
  \centerline{\includegraphics[scale=0.3]{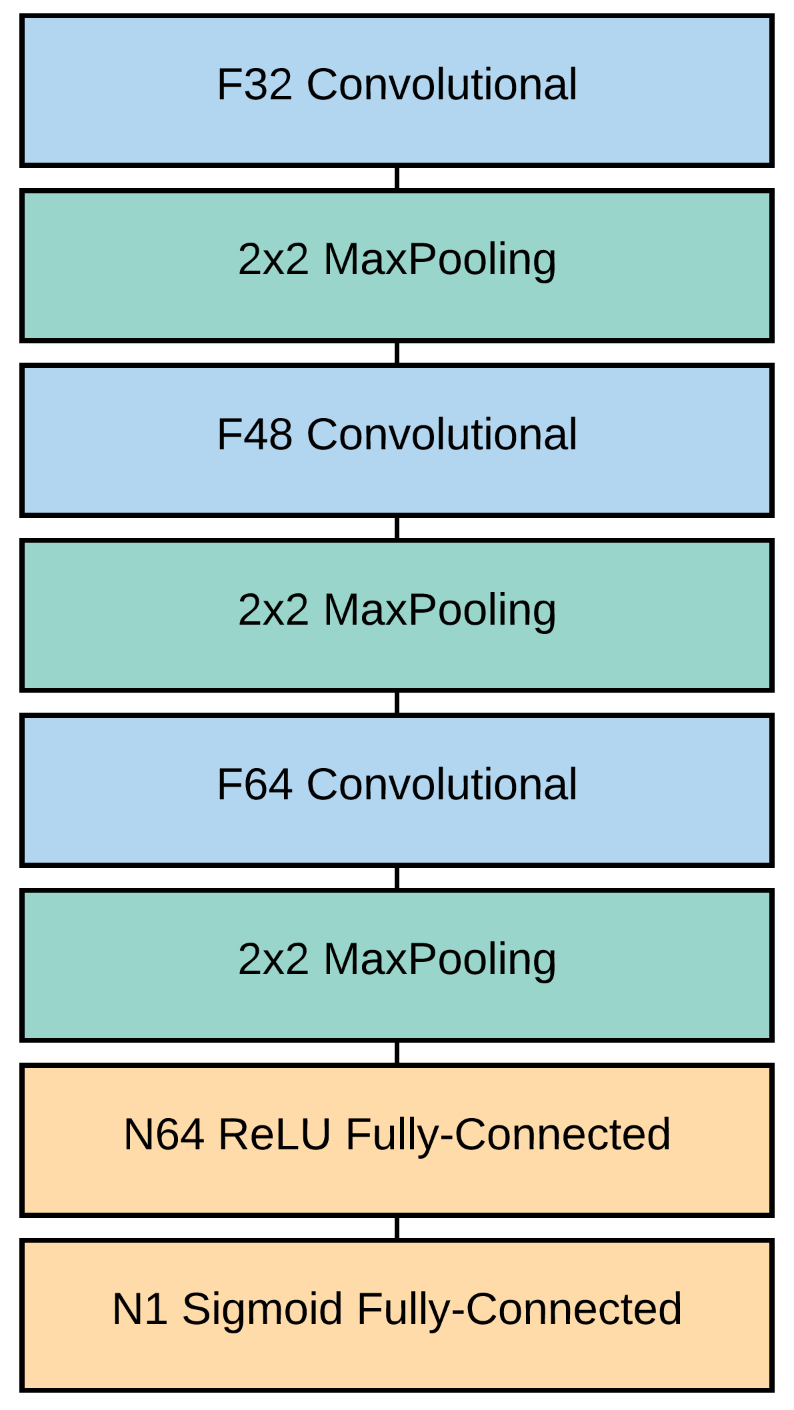}}
  \caption{Network architecture for single CNN. Input image (tensor) at top. Output prediction at bottom.}
  \label{single_architecture}
\end{figure}

We initally choose the architecture by \cite{Chollet2016}, a relatively modest network architecture with (only) 3,714,593 free parameters (see Figure \ref{single_architecture}) designed for smaller datasets. Convolutional layers have Fn (e.g. F32) 3x3 convolutional matrices (i.e. filters), each with a convolution step size (i.e. stride) of 1x1. Fully-connected layers have Nn (e.g. N64) neurons. The final layer is a single neuron whose output represents the continuous-valued class prediction.

We verify with a grid search that this architecture outperforms three convolutional layer networks with significantly higher or lower numbers of convolutional filters or layers. Three convolutional layers provide enough depth for high performance without becoming computationally intractable, while the relatively low number of filters helps prevent overfitting. The majority of free parameters (3,686,464) are in the first fully-connected layer.  

To further minimise overfitting, we apply `dropout' to this layer. Intuitively, dropout temporarily removes random selections of neurons. This encourages neurons to learn parameters that remain discriminative for many different combinations of other neurons in the network. A neuron and all associated connections are referred to as a unit. For each training epoch, each unit in the fully-connected network layer has probability $p$ to be removed for that epoch. The operation of a fully-connected layer with dropout is

\begin{equation}
x_{i}^{(l)} = f(\mathbf{w}_{i}^{(l)}\mathbf{x'}^{(l-1)} + \mathbf{b}_{i}^{(l)})
\end{equation}

where $\mathbf{w}_{i}^{l}$ denotes unit $i$ in layer $l$, $x_{i}^{(l)}$ is the (scalar) output of unit $i$ in layer $l$ and $\mathbf{x'}$ is the elementwise product
$\mathbf{x'} = \mathbf{x} \text{ } (*) \text{ } \mathbf{B}(p)$ with $\mathbf{B}(p)$ being an $\mathbf{x}$-shaped matrix with binary elements according to the Bernoulli distribution (i.e. 1 with probability $p$, 0 with probability $1-p$). 

The thinned network (following dropout) is trained for a single epoch before $\mathbf{B}(p)$ is re-evaluated, causing different units to be active and a new thinned network to be created (sharing weights with the predecessor). Dropout therefore effectively trains many unique networks, increasing prediction quality. We select the hyperparameter $p$ to be 0.5, based on the results of the grid search described below.  

\subsection{Augmentation}
\label{augmentation}

Galaxy morphological classifications should be invariant under certain transformations, such as  flips, rotations, minor zooms, and minor translations. CNNs lack our intuitive understanding of transform invariance and require sufficiently diverse examples to infer which transforms are not discriminative.  We therefore artificially expand our training set by including many variants of the original input images. By inputting many randomly-transformed images with unchanged labels, we teach the network to be insensitive to those transforms.  By applying these transforms dynamically when each input image is read by the network, the effective training set becomes arbitrarily large and the network always sees a unique image.  
Note that augmented images are less informative than truly new images; once the network has learned the invariance, further augmented images do not improve performance.

We randomly apply each of the following transforms to augment the images:

\begin{enumerate}
\item Horizontal flip
\item Vertical flip
\item $\pm \frac{\pi}{2}$ resampled rotation
\item 90\% to 110\% resampled zoom
\item $\pm 5\%$ horizontal translation
\item $\pm 5\%$ vertical translation
\end{enumerate}

To avoid unnecessary information loss from pixel resampling after each step, the transforms are applied through a single net transformation. We verify with a grid search that the resolution degradation from the net transformation has a less significant impact on prediction quality than the corresponding learned invariance. 

\begin{figure}
  \includegraphics[width=\columnwidth]{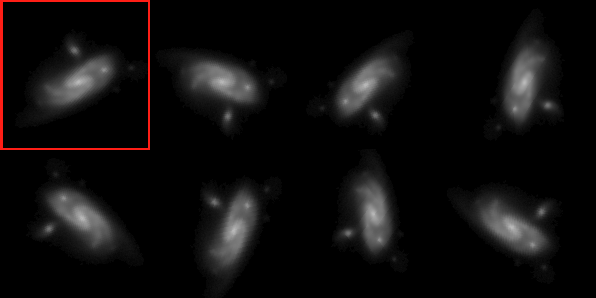}
  \caption{Mosaic of illustrative augmented images of a single non-tidal galaxy. Images are mean-averaged, masked (3$\sigma$) and shown in log scale. The images are cropped from 256 to 150 pixels for illustration only. The original image is shown in the top left (red highlight).}
  \label{augmented_images}
\end{figure}

Figure \ref{augmented_images} shows a single galaxy with seven different augmentations applied. The same random augmentation process creates a unique image each time. 

\subsection{Grid Search}
\label{gridsearch}

CNNs have tuneable design values (e.g $\text{layer count} = 4$, $\text{first layer width} = 256$) called hyperparameters \citep{Lu2015}. The choice of hyperparameters may have a significant impact on classifier performance, but the optimal choice is not known {\it a priori}.  Estimates can be made with heuristics (rule-of-thumb guesses) based on previous generic image classification work. However, images of galaxies with faint tidal features are unlike `terrestrial' pictures in that they have extreme contrast, high noise levels and indistinct subject shapes, and so borrowing from such work is unlikely to be optimal. 

We improve our hyperparameter estimates using grid searches. Through this procedure, many possible network configurations are trained and measured. We choose to separate hyperparameters into related groups and then identify the optimal choice within that group through an exhaustive grid search. For example, we assume the optimal number of layers is independent of pixel rescaling and proceed to test many possible numbers of layers with a single rescaling. This approach makes the grid search computationally feasible without needing to specify any hyperparameters with heuristics.  

We use three groups of hyperparameters: preprocessing (see Section \ref{preprocessing}), architecture (see Section \ref{network_architecture}), and augmentation (see Section \ref{augmentation}). We find the best performing preprocessing configuration is band-stacked images with $3\sigma$ masking,  logarithmic pixel intensity scaling and no mean convolutions. Performance is invariant under physically reasonable choices of pixel clipping values and,  provided that mean convolutions are not used, also invariant under pixel intensity rescaling.  This latter result is intuitively surprising given the impact that rescaling has on human perception. The best performing architecture and augmentation configurations have already been discussed in Sections \ref{network_architecture} and \ref{augmentation}, respectively.

\subsection{Single Network Results}
\label{singleresults}

We select completeness and contamination as metrics to evaluate the performance of our network. Conceptually, completeness is the probability for a visually-classified tidal galaxy to be correctly identified by the CNN as tidal, and contamination is the probability that a visually-classified non-tidal galaxy is incorrectly identified by the CNN as tidal. Mathematically, these are the true positive rate (TPR) and false positive rate (FPR), respectively:
\begin{equation}
TPR=\frac{TP}{TP+FN} \hspace{1cm} FPR = \frac{FP}{FP+TN}.
\end{equation}

\begin{figure}
  \includegraphics[width=\columnwidth]{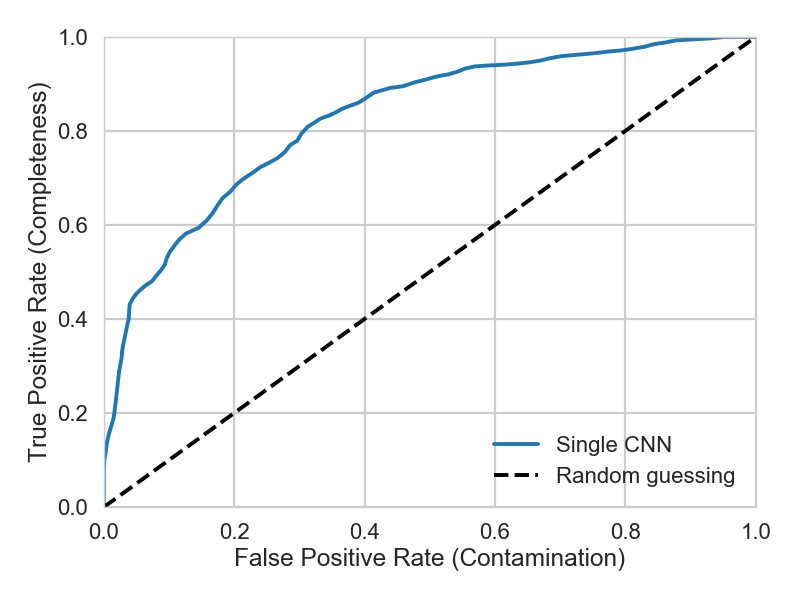}
  \caption{The ROC curve for a single CNN classifier on CFHTLS-Wide images. The dashed line indicates the expectation for random guesses.}
  \label{single_roc}
\end{figure}

The Receiver Operating Characteristic (ROC) curve illustrates the completeness and contamination of the classifications. The ROC curve of our best-performing single CNN classifier is plotted in Figure \ref{single_roc}. The completeness and contamination may be selected along any point on the curve, corresponding to varying the confidence threshold used to classify images as tidal. For example, one might choose a completeness of 70\% and therefore a contamination of 22\%. Random guessing would provide equal completeness and contamination.

\begin{figure}
  \includegraphics[width=\columnwidth]{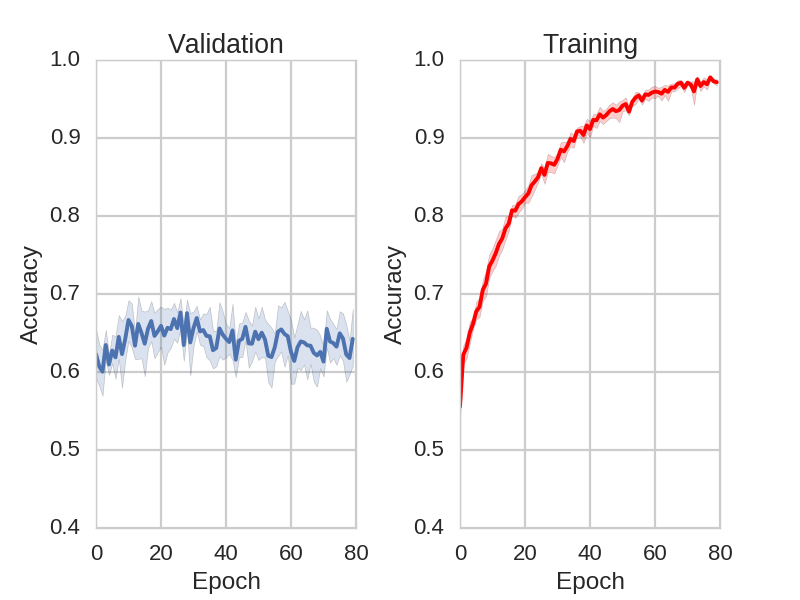}
  \includegraphics[width=\columnwidth]{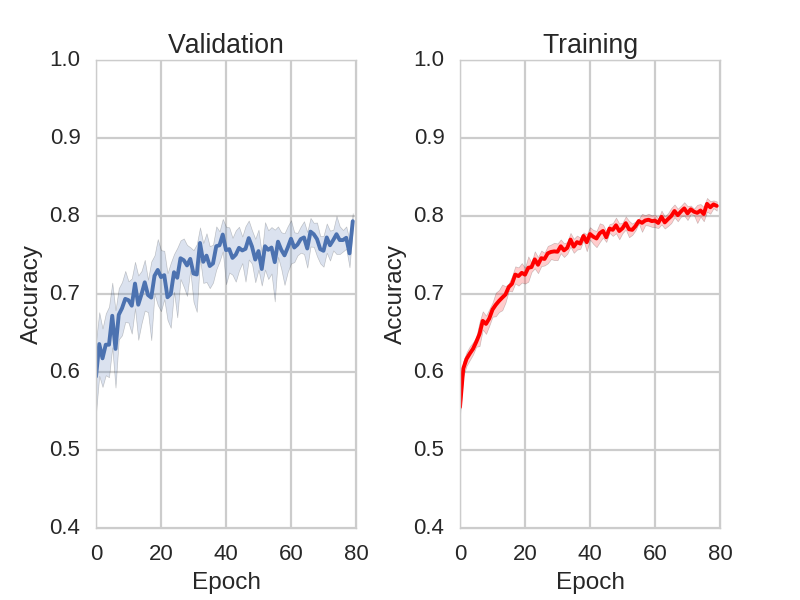}
  \caption{Mean training metrics for the same network architecture with dropout and augmentations off (top) vs. on (bottom), over five runs (trained and validated on each five-fold cross-validation permutation). Shaded regions denote the 90\% Bayesian credible interval.}
  \label{dropout_result}
\end{figure}

Figure \ref{dropout_result} shows the accuracy of a single classifier with and without dropout and augmentations, averaged over five runs. Shaded regions denote the 90\% Bayesian credible interval \citep{Oliphant2006}. Without dropout and augmentations, the training accuracy increases with the number of galaxies that the network sees while the validation accuracy remains low. This is because the network is overfitting to random patterns in the training data. These patterns do not generalise to new data so the validation accuracy remains low. In contrast, with dropout and augmentations, the network is learning patterns present in both the training and validation data, causing both accuracies to rise. 

\begin{figure}
  \includegraphics[width=\columnwidth]{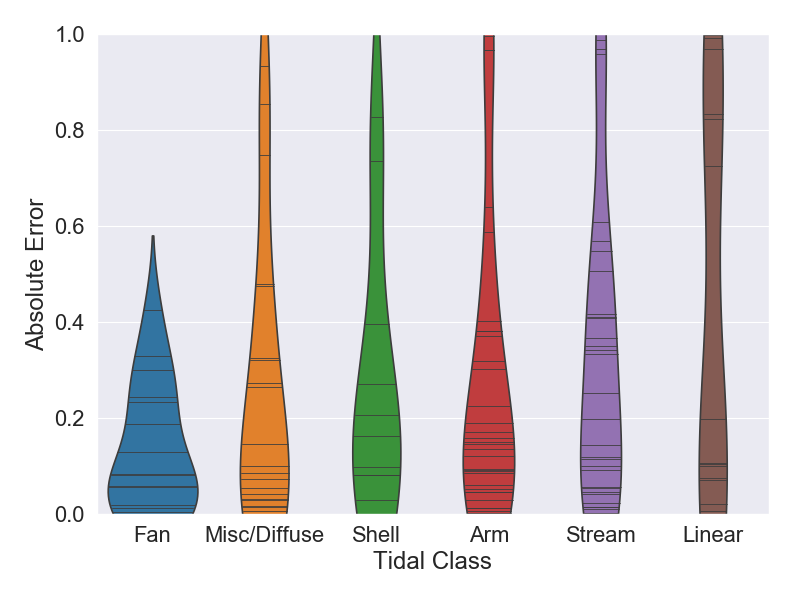}
  \caption{Single network validation performance by class of tidal feature. 
  Each column is a class of tidal feature, ordered left to right by increasing mean absolute error on galaxies with that feature. 
  Horizontal black bars denote individual galaxies: for example, a galaxy with a fan feature on which the network prediction had an absolute error of 0.2.
  More galaxies with lower absolute error indicates better performance.
  The area of each column illustrates the probability density, inferred (by kernel density estimate) from all galaxies of that class.
  Feature classes follow the schema introduced by A13.}
  \label{performance_by_class}
\end{figure}

Figure \ref{performance_by_class} shows performance broken down by class of tidal feature, following the schema introduced by A13.
Every prediction is made by a network which has not been trained on that galaxy, following the cross-validation strategy described in Section \ref{training}.
Networks perform best (i.e. have the lowest mean absolute error) on fan features, a surprising result given the relative rarity of such features. 
In general, performance is higher for dispersed features (fan, diffuse, shell) than small-scale structural features (arm, stream, linear).
We speculate that this may be because such features are unlikely to be mimicked by contaminant objects in the field-of-view, and therefore easier to learn from our relatively small dataset.

\begin{figure*}
  \begin{subfigure}[b]{0.49\textwidth}
      \includegraphics[scale=.35]{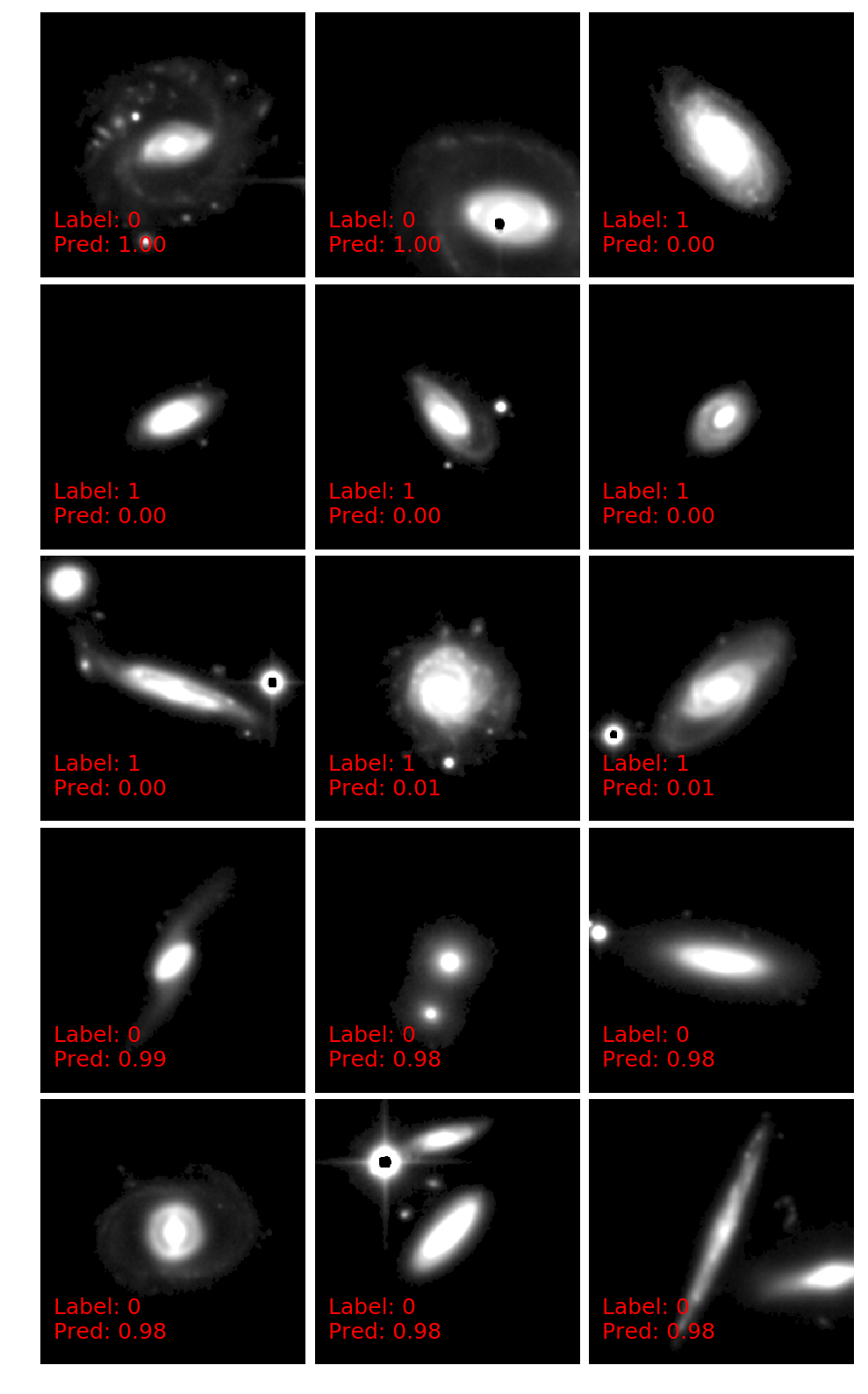}
  \end{subfigure}
  \hfill
  \begin{subfigure}[b]{0.49\textwidth}
      \includegraphics[scale=.35]{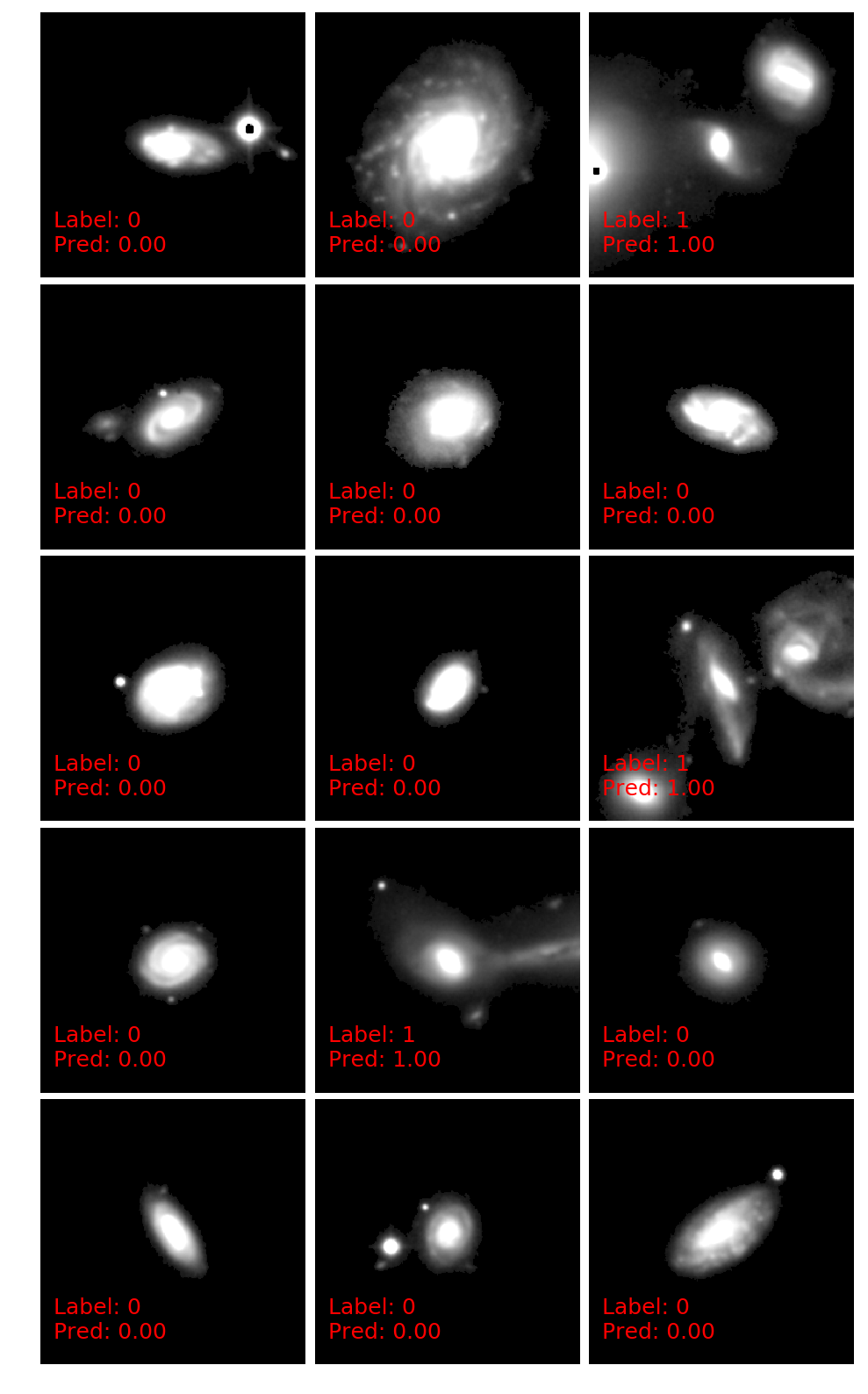}
  \end{subfigure}
  \caption{Galaxies with the highest (left) and lowest (right) absolute error in validation predictions, as presented to the network following the preprocessing strategy identified as optimal (including pixel rescaling and background masking, see Section \ref{preprocessing}). Brightness and contrast have been further adjusted for human viewing of tidal features.}
  \label{failure_examples}
\end{figure*}

All classes except fan (which is both rare and has a low mean error) have at least one prediction with an error close to one.
This reflects the probabilistic nature of the method; statistical metrics of success do not imply that every prediction is approximately correct.
Figure \ref{failure_examples} shows the galaxies with the highest and lowest absolute error (matching the highest and lowest horizontal bars across all columns in Figure \ref{performance_by_class}).
Failures show no obvious pattern, underscoring how the operation of convolutional neural networks is not always immediately interpretable by humans. 
We investigate the behaviour of the network in Section \ref{heatmap}.

\section{Ensemble Classifier}
\label{ensemble}

\subsection{Configurations}
\label{configurations}

The predictions of an ensemble of classifiers will typically outperform those of a single classifier because each independent prediction provides new information on the classification of the input image. This information is exploited through averaging over the individual predictions (e.g. \citealt{Zhang2012}). Indeed, ensemble methods are routinely used to improve image classification performance e.g. \citep{Dieleman2015}.  We investigate two different ensemble configurations -- CNN using optimal preprocessing (configuration A), and CNN using varied preprocessing (configuration B) --  as a means to generate more accurate faint tidal feature classifications for our sample.

In configuration A, each CNN is in the optimal hyperparameter configuration identified in Section \ref{single}. The random order of input training images and the random initialisation of weights and bias prior to training may cause the CNN to converge to different local minima during training, particularly when applied to smaller training sets \citep{LeCun2015}. This leads to identically-configured CNNs making slightly different predictions, which is described as stochastic independence.

In configuration B, each CNN uses varied preprocessing hyperparameters, as detailed below. This introduces further independence between CNNs. Different preprocessing hyperparameters might lead a CNN to advantageously detect different tidal features. For example, more restrictive masking thresholds will reduce the number of contaminant objects in the field-of-view but may also reduce the spatial extent of particularly faint tidal features.  However, hyperparameters that are different to the optimal hyperparameters will degrade the performance of a single model. By comparing each ensemble configuration, we test if (for our problem) it is more effective to ensemble individually stronger classifiers with lower independence (configuration A) or individually weaker classifiers with higher independence (configuration B). 

We select the following set of preprocessing hyperparameters for the five CNNs comprising the configuration B:

\begin{enumerate}
\item Logarithmic rescaling, $3\sigma$ mask threshold (i.e. optimal)
\item Logarithmic rescaling, $5\sigma$ mask threshold
\item No (linear) rescaling, $3\sigma$ mask threshold
\item No (linear) rescaling, $5\sigma$ mask threshold
\item No (linear) rescaling, band-stacked (un-masked) image
\end{enumerate}

These were chosen for being high-performing combinations identified with the grid search described in Section \ref{gridsearch}, and for spanning visually distinct preprocessing steps.

\subsection{Training and Evaluation}

\begin{figure*}
  \includegraphics[scale=0.8]{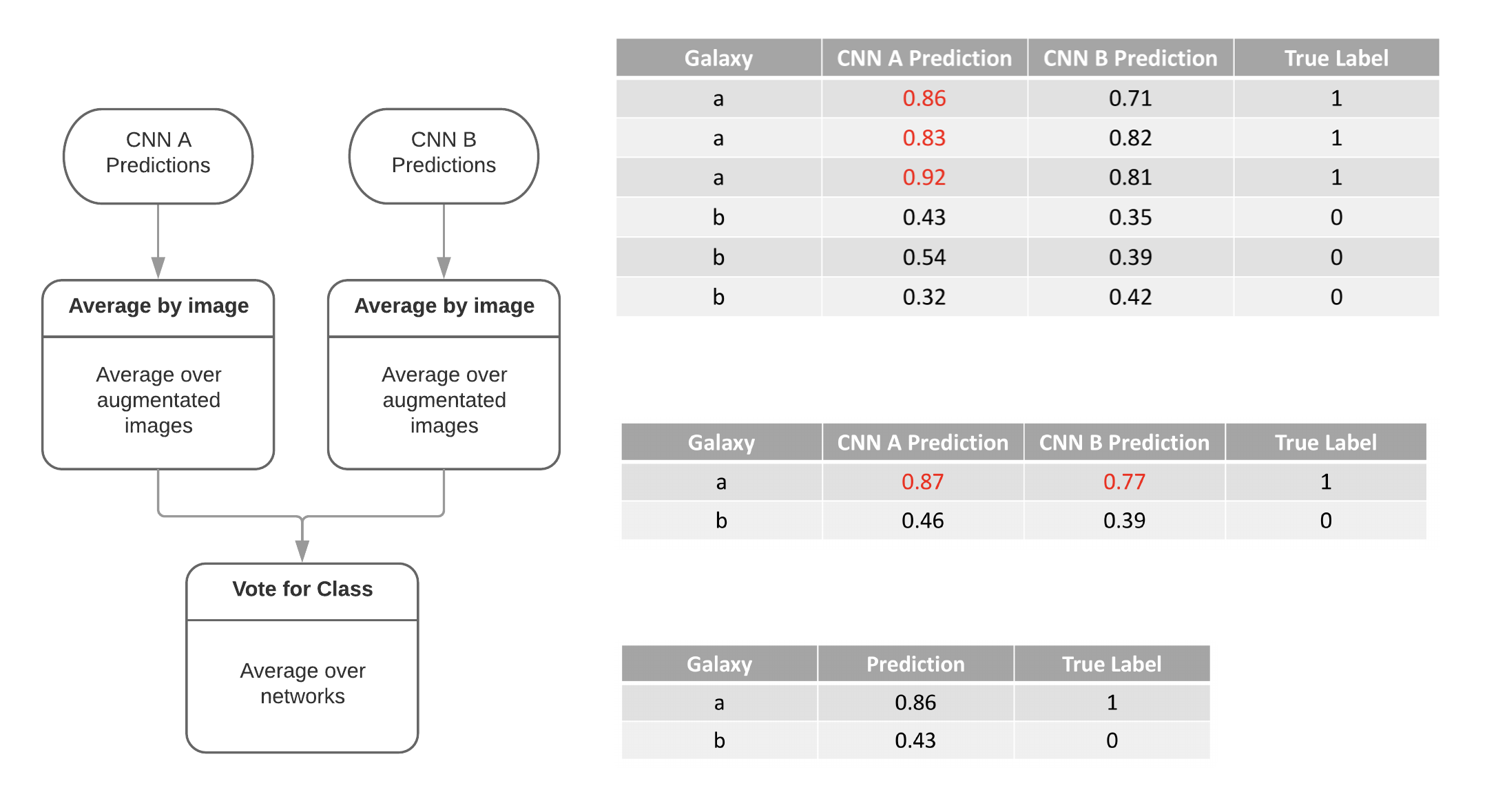}
  \caption{Flow chart of each stage for ensemble classifier. Red text illustrates the values being combined at each stage.}
  \label{ensemble_flow}
\end{figure*}

To decide which configuration has the best performance, we need to train and evaluate all five CNN composing that configuration.  Each CNN is trained on images that are randomly drawn in equal measure from 80\% of the tidal and non-tidal classes, as described in Section \ref{training}. Each CNN is then asked to make several predictions on each galaxy in the remaining `unseen' 20\%. We then calculate an overall prediction for the configuration by combining the predictions of each CNN. Figure \ref{ensemble_flow} illustrates how the predictions of each CNN are combined. 

First, for each CNN, we average over all predictions made by that CNN on augmented images of the same galaxy. We know that the true label is invariant under our augmentations  but the CNN may not have completely learned to ignore them. Averaging over predictions of the same galaxy ensures that our final configuration prediction will not depend on any particular augmentation.

After recording the augmentation-averaged prediction on each galaxy by all five independently-trained CNN, we then average those single-CNN predictions which allows us to exploit any independence in those predictions to improve performance, as explained in Section \ref{configurations}.

\subsection{Results}
\label{ensemble_results}

\begin{figure}
  \includegraphics[width=\columnwidth]{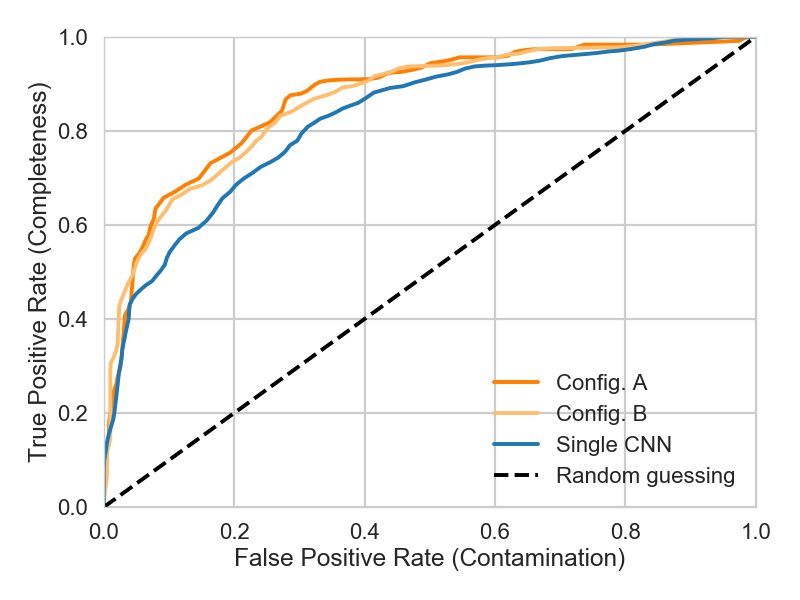}
  \caption{Comparison of the ROC curves of a single CNN, and our two ensemble CNNs.}
  \label{ensemble_roc}
\end{figure}

Figure \ref{ensemble_roc} shows the average ROC for each ensemble configuration, and overplots the ROC of the individual optimal CNN shown in Figure \ref{single_roc}. 
We find that both ensemble configurations provide a significant improvement over using a single optimal CNN, as expected. 
For example, with a single CNN we were able to achieve a completeness of 70\% with a contamination of 22\%. 
This now improves to a completeness of 76$\% \pm 2\%$ with the same level of contamination. 
The prediction quality of the two ensemble configurations is approximately equal within the expected statistical variation. 
That is, for our problem, both configurations are equally effective.

The ROC curve measures performance when classifying all galaxies, which is appropriate for understanding the overall performance of a method.
In practice, we might instead choose to classify only a subset of galaxies for which the model is reasonably confident, and refer the remainder to experts or citizen scientists.
We can measure model confidence using the continous prediction score output by the model.
By optimising our model using the binary cross-entropy loss (Equation \ref{loss}), which heavily penalises mistaken scores near 0 or 1, we can interpret scores near 0 or 1 as confident predictions and scores close to 0.5 as uncertain predictions \citep{Tewari2007}.
Therefore, we can select galaxies with confident predictions by requiring a score at least some minimum difference from 0.5.

However, because the model was trained on an equal number of tidal and non-tidal galaxies (Section \ref{training}), our scores on the full imbalanced sample are uncalibrated; the model doesn't know that non-tidal galaxies are common.
To account for this, we calibrate our scores with Platt's Scaling \citep{Fonseca2017}. 
We use logistic regression to fit a correction to the fraction of true positives on 25\% of galaxies, such that the scores match the empirical probability that a galaxy is tidal, and then apply that correction to the scores of the remaining galaxies.

\begin{figure}
  \includegraphics[width=\columnwidth]{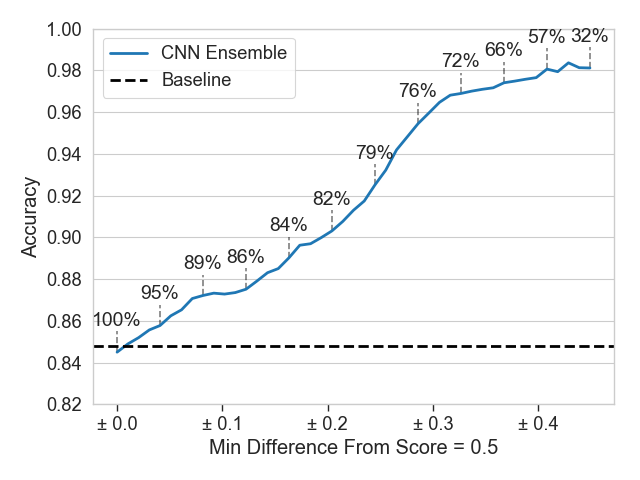}
  \caption{Accuracy of CNN ensemble (Config. A) on subsamples with increasingly confident predictions. 
  Accuracy is measured for galaxies where the classifier score is a given minimum difference from 0.5. 
  The greater the minimum difference, the more confident the classifier is.
  The percentage of galaxies with at least that confidence is annotated.
  For example, 72\% of galaxies have a min. score difference of at least $\pm 0.33$ (i.e. a score above 0.83 or below 0.17) and can be classified with 97\% accuracy.
  Also shown is a baseline classifier that always predicts non-tidal (the majority class).
  }
  \label{accuracy_by_confidence}
\end{figure}

Having calibrated our scores, we can now measure how performance varies on increasingly confident subsamples.
We find that performance can be dramatically improved, at the cost of leaving some galaxies unlabelled.
Figure \ref{accuracy_by_confidence} shows how the accuracy varies as we classify only galaxies where the model is increasingly confident.
On the full sample, the calibration causes the model to predict `non-tidal' on three out of four galaxies, leading to an accuracy similar to a baseline classifier that always predicts non-tidal.
However, by using the score to identify galaxies where the model is more confident, we can make useful predictions on the bulk of the sample.
For example, the 72\% of galaxies with a min. score difference of at least $\pm 0.33$ can be classified with 97\% accuracy, compared to 84\% accuracy on all galaxies.
This suggests that our prototype model can be used to identify the bulk of a survey with near-perfect accuracy, drastically reducing the human labelling effort required to create extensive science-ready catalogues of galaxies with or without tidal features.

We next investigate the independence of the single classifiers within each ensemble by measuring the correlation between each possible pair of classifiers. 

\begin{figure}
  \includegraphics[width=\columnwidth]{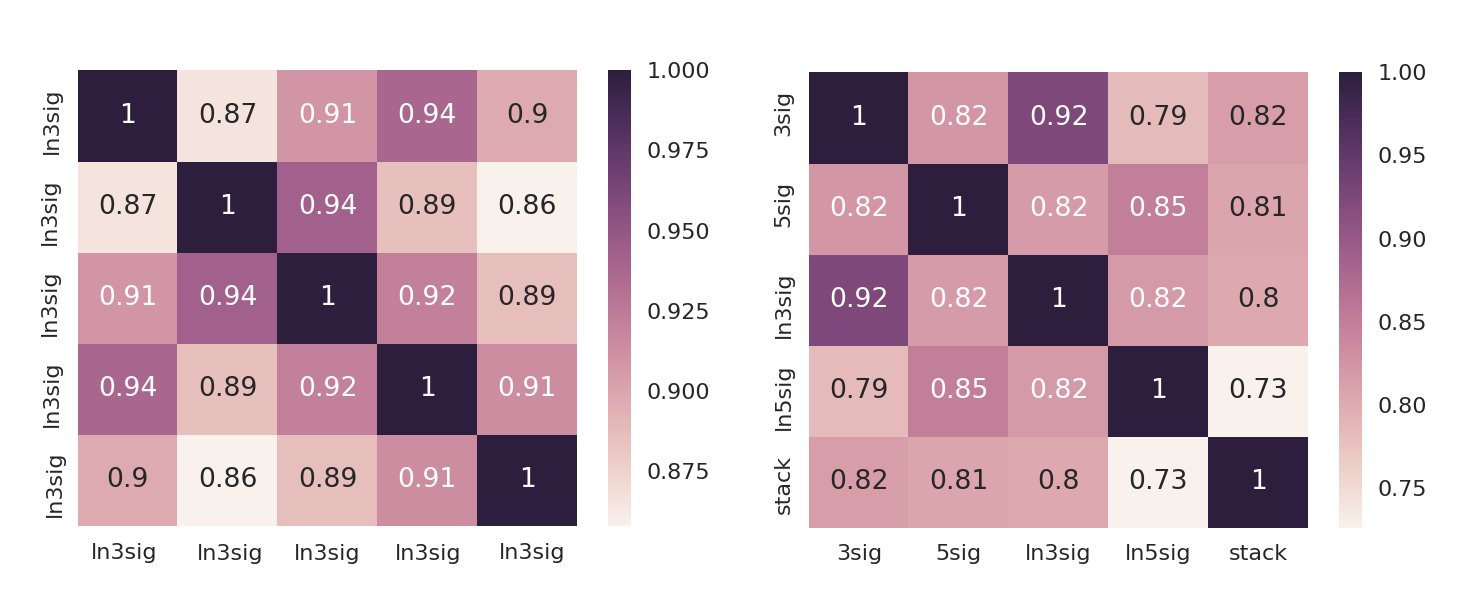}
  \caption{Pearson correlation coefficients between the predictions of single CNN (rows, columns) classifiers acting within ensembles using optimal preprocessing (A, left) and varied preprocessing (B, right). Labels denote the preprocessing used for that CNN, with `ln' denoting logarithmic pixel rescaling and `Nsig' denoting the masking threshold used. Configuration A combines five classifiers all using the same optimal preprocessing configuration (logarithmic rescaling and a 3-sigma pixel mask) while configuration B combines five classifiers with varied preprocessing configurations.}
  \label{correlation}
\end{figure}

The correlation is measured with the Pearson $r$ correlation coefficients between the continuous-valued predictions of each classifier. 
The resulting matrices are shown in Figure \ref{correlation} and are symmetric due to the symmetry of the correlation coefficient. 
Unitary diagonal elements result from pairwise comparisons between a CNN and itself, and may therefore be neglected. 

Recall that configuration A combines five classifiers all using the same optimal preprocessing configuration (logarithmic rescaling and a 3-sigma pixel mask, see Section \ref{gridsearch}), while configuration B combines five classifiers with varied preprocessing configurations.
The average (non-unitary) correlation coefficient is lower for the ensemble with varied preprocessing (B, $\bar{r}$=0.82) than with optimal preprocessing (A, $\bar{r}$=0.90), indicating that \emph{additional independence can be introduced by altering the preprocessing process}.
In particular, altering the masking threshold has a greater effect on classifier predictions than changing from logarithmic to linear rescaling. This is consistent with our earlier finding that prediction accuracy is invariant within statistical uncertainty under pixel rescaling.

\section{Comparison with Current Methods}
\label{comparison_with_current_methods}
As discussed in Section \ref{introduction}, most current methods of automated feature detection are not well-suited to recovering the typical low surface brightness tidal features that arise from minor mergers and accretions.  We have selected two promising alternative methods from the recent literature and applied them to the A13 sample in order to benchmark their performance against that of the CNNs. These are:

\begin{enumerate}
\item Shape asymmetry \citep{Pawlik2016}, an example of a method based on non-parametric feature extraction; 
\item WND-CHARM \citep{Shamir2012,Schutter2015}, an alternative unsupervised machine learning approach previously shown to be successful in identifying peculiar and interacting galaxies.
\end{enumerate}

Detecting tidal features by any method is dependent on:
 
\begin{enumerate}
\item The nature of the sample under study. The varying depths, bandpasses and spatial resolutions of different datasets can lead to incomparable detection rates; 
\item The author's definition of what is tidal. The context of the paper often sets the definition for a tidal feature, and different authors may reasonably have different definitions.   
\end{enumerate}

For example, \cite{Bridge2010} and A13 both use data from the CFHTLS to identify tidal features through visual inspection. However, \citealt{Bridge2010} uses data from the Deep component of the survey, which covers less sky area but is sensitive to more distant galaxies than the Wide component used by A13. Furthermore, they select different features to define which galaxies are tidal (tidal tails and bridges vs. the more subtle debris features outlined by A13). Directly comparing the detection rates (and underlying methodology) of these two papers is therefore not meaningful as they measure different things. 

Through applying all three methods to the same galaxy sample, with the same binary labels, we sidestep many of the complications that arise when comparing results that have appeared in the literature. We also ensure that the ability of each classifier to detect {\it the same} tidal features is tested fairly. Below, we describe each method and motivate why we have selected that particular method for comparison.

\subsection{Application of the Shape Asymmetry method}
\label{pawlik}
\label{pawlik_comparison}

Shape asymmetry was introduced by \cite{Pawlik2016} as a method to automatically detect faint asymmetric tidal features in galaxies that experienced a recent merger. It is an appropriate choice for tidal feature detection in galaxies with complex morphologies since, unlike residual-based methods, it does not require a parametric fit of the underlying galaxy light profile.  The measure is only sensitive to morphological asymmetry and does not contain information about the asymmetry of the light distribution. When applied to a sample of 70 starburst and post-starburst galaxies imaged by the Sloan Digital Sky Survey \citep{Abazajian2009}, \cite{Pawlik2016} report an accuracy of 95\% in detecting post-merger tidal features.

The method works as follows. First, following \cite{Conselice2003} and \cite{Conselice2014}, the minimal asymmetry centroid is identified and asymmetry parameter $A$ is recorded. 

\begin{equation}
A = \frac{\sum{\mid I_0 - I_{180} \mid}}{2 \sum{\mid I_{180} \mid}} - A_{bgr}
\end{equation}
where $I_0$ is the value of a galaxy pixel, $I_{180}$ is the value of the pixel at the same position after the image is rotated $180\deg$, $A_{bgr}$ is the estimated contribution to asymmetry from background noise, and all sums act over all pixels. Note that low surface brightness pixels will have small $I_0$ and hence will make only minimal contributions to $A$. As a result, $A$ is relatively insensitive to faint tidal features. 

\begin{figure}
  \includegraphics[width=\linewidth]{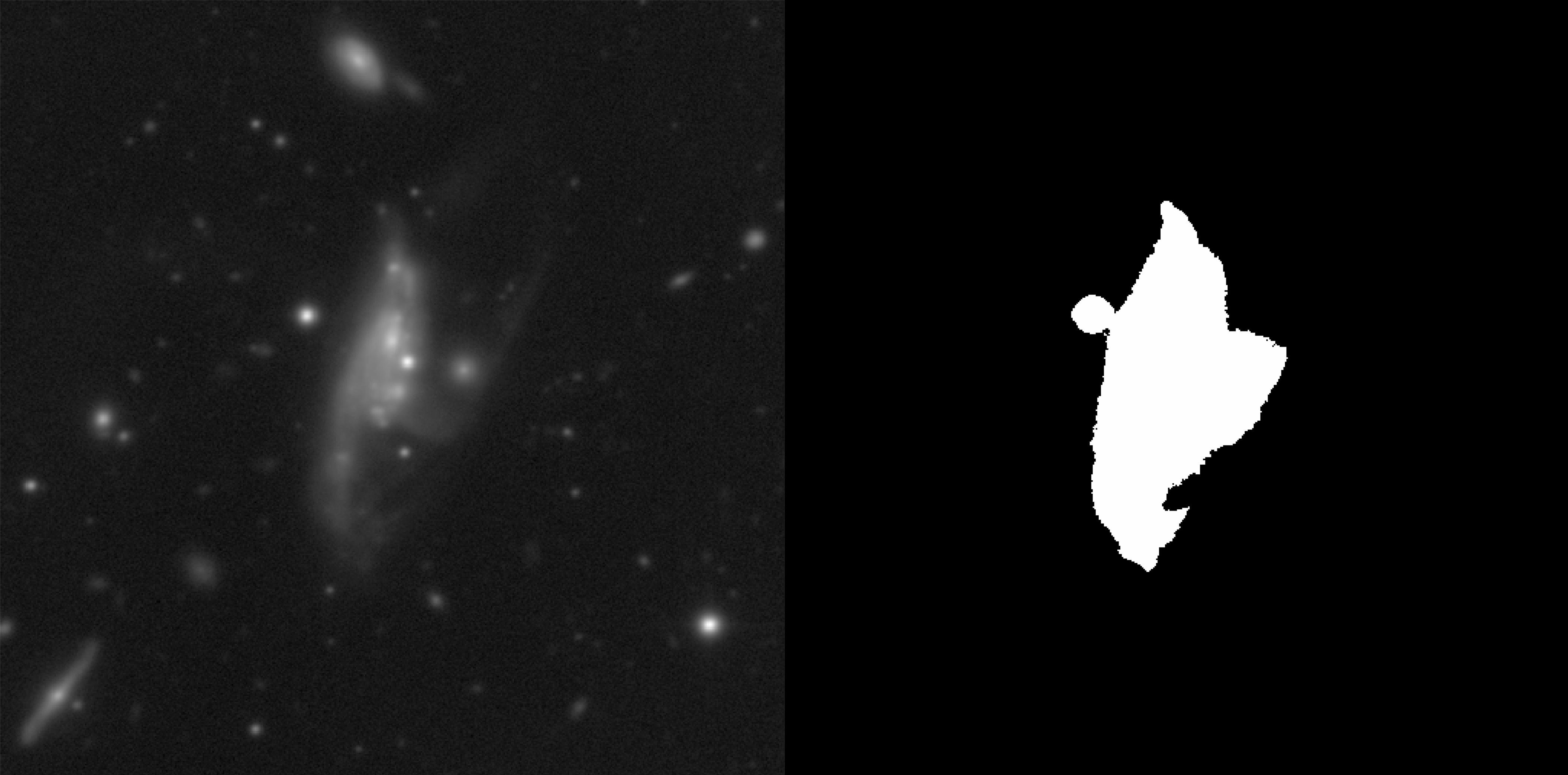}
  \caption{Illustration of the non-parametric shape asymmetry method of Pawlik et al 2016, applied to a galaxy in the CFHTLS-Wide sample. Left: stacked \textit{gri} galaxy image (logarithmically rescaled for illustration only). Right: binary mask of pixels above $3\sigma$ used to calculate shape asymmetry $A_{s}$.}
  \label{pawlik_method}
\end{figure}

Next, a $3$x$3$ mean convolution is applied to the galaxy image to enhance low surface brightness features. A binary mask is then created with values of 1 where the corresponding pixel count is both some chosen N$\sigma$ above the original measured sky background and contiguously eight-connected to the central pixel. The intuitive effect is to create a silhouette of the galaxy outline that includes faint structure - see Fig \ref{pawlik_method}. For our implementation, background estimation is done with the procedure described in Section \ref{preprocessing}. We find a pixel masking threshold of $N=3$ gives optimal results.  

Finally,  the shape asymmetry parameter $\text{A}_{s}$ is calculated in analogy to $A$ but with $I_0$ and $I_{180}$. replaced by the pixel values of the binary mask, rather than the original galaxy image:

\begin{equation}
A_s = \frac{\sum{\mid M_0 - M_{180} \mid}}{2 \sum{\mid M_{180} \mid}}
\end{equation}
where $M$ ($M_{180}$) is the value of a mask pixel at some (rotated $180 \deg$) position on the binary mask. 

To ensure tidal features at the image extremities are included, the selection radius used to calculate both $A$ and $\text{A}_{s}$ is taken as the minimum radius that encloses the full binary mask.  By plotting $A$ against $A_s$, an empirical selection cut can be made to identify galaxies with tidal features. 

\begin{figure}
  \includegraphics[width=\columnwidth]{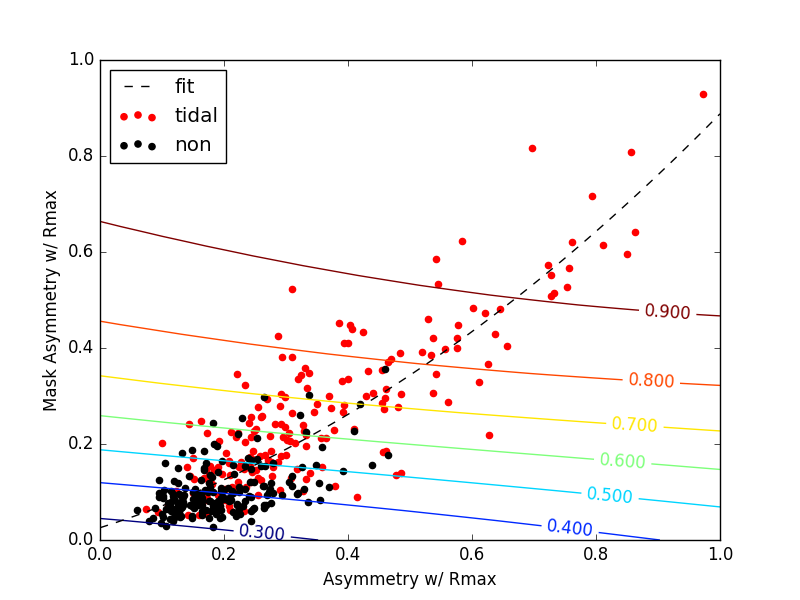}
  \caption{Probability space generated by Pawlik method on 500 CFHTLS-Wide galaxies from the A13 sample, illustrated by contours. Mask asymmetry is the shape asymmetry $A_{s}$. Galaxies are observed to follow a clear linear trend on the mask asymmetry/asymmetry space, which we fit for interest only.}
  \label{pawlik_probability}
\end{figure}

Figure \ref{pawlik_probability} shows the resulting asymmetry space for our CFHTLS-Wide sample where 250 random examples are plotted per binary class. On the basis of visual inspection, \cite{Pawlik2016} chose an empirical cut of $A_{s} > 0.2$ to select tidal galaxies. However, we would prefer to understand how the shape asymmetry method balances completeness and contamination. To do this, we generate ROC curves using two methods (each generating a slightly different curve).  In the first method, we generalize the $A_{s} > 0.2$ sample cut by making many sample cuts separated by $\delta A_{s}$. The ROC curve is then calculated in the continuous limit $\delta A \xrightarrow{} 0$.   In the second method, we divide the galaxies into five subsets, train a logistic regression classifier \citep{Pedregosa2012} implemented in \texttt{scikit-learn}  on four subsets, and make predictions on the remaining test partition. This is repeated for each combination of partitions (i.e. five-fold cross-validation). The ROC curve is calculated as the mean ROC curve over the test predictions for combination. To verify that the logisitic regression classifiers are functioning correctly, we illustrate the mean estimated tidal probabilities at every point with contours on Figure \ref{pawlik_probability}.

\subsection{Application of the WND-CHARM algorithm}
\label{wndcharm_comparison}

Here we explore the general-purpose image classification algorithm WND-CHARM \citep{Orlov2008}. It provides an example of unsupervised machine learning and is publicly available as both a command-line tool and Python API from \url{https://github.com/wnd-charm/wnd-charm}. 

Like CNNs, WND-CHARM was originally developed for other uses \citep{Orlov2008} and was only later applied to astronomy.  It has been successfully used to classify peculiar galaxies \citep{Shamir2012} and general galaxy morphology
\citep{Schutter2015} and so could be reasonably expected to perform well on the problem of faint tidal debris. WND-CHARM represents a middle ground between algorithms that use user-defined image features (for example, random forests - see \citealt{Ball2010}) and algorithms that infer the ideal features to construct from pixel data (for example, CNNs).

WND-CHARM \citep{Orlov2008} identifies the macroscopic properties (e.g. total image brightness) which are most discriminative between classes in the training sample. Those properties are then used to classify test images by identifying the class of the closest known example, with `closest' defined in a multidimensional Euclidean space where each dimension corresponds to a discriminative property. 

The augmentation procedure we use for our convolutional network (see Section \ref{augmentation}) is designed to improve classifier performance. To provide a fair comparison, we train and test WND-CHARM on subsets of 25,000 images preprocessed and augmented through the same procedure.  We use a train-test split of 80\% and 20\% respectively to evaluate the algorithm.  

\subsection{Overall Comparison}
\label{overall_comparison}

\begin{figure}
  \includegraphics[width=\columnwidth]{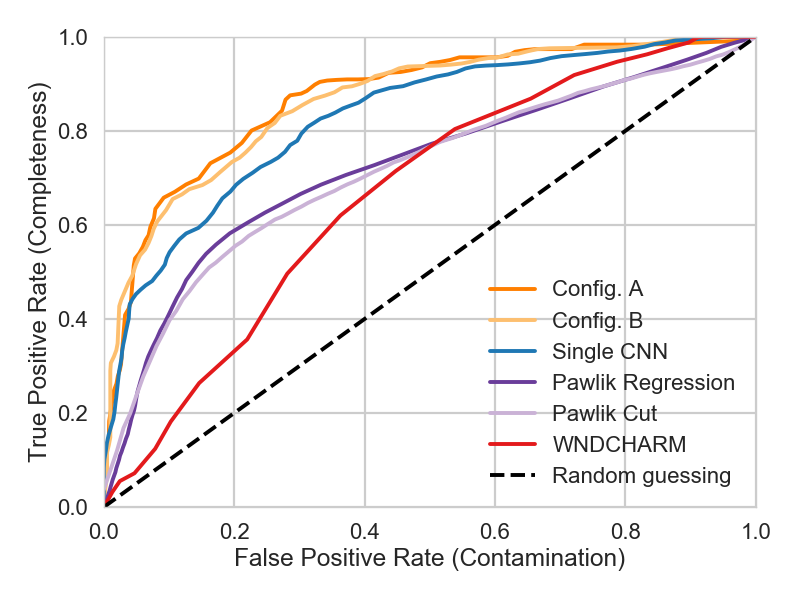}
  \caption{The ROC curves for all classifiers tested on the A13 sample.}
  \label{allrocs}
\end{figure}

\begin{figure}
  \includegraphics[width=\columnwidth]{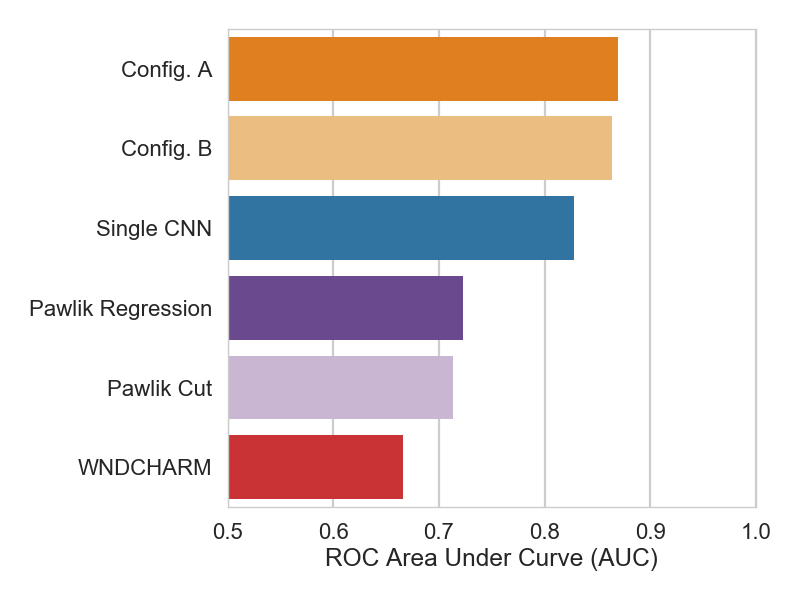}
  \caption{The ROC area-under-curve (AUC) values for all classifiers tested on the A13 sample.}
  \label{allaucs}
\end{figure}

Figure \ref{allrocs} shows the completeness and contamination achieved by the three approaches over many confidence thresholds (not shown). Figure \ref{allaucs} summarises overall performance with the area-under-curve (AUC) scores for each method. The AUC score is frequently used in machine learning literature as a scalar summary metric of classifier quality \citep{Huang2005}. For our problem, the AUC score measures the probability that a random galaxy with tidal features will correctly be recognised as being more likely to have such features than another random galaxy without tidal features. 

It is readily apparent that our CNNs have higher completeness and lower contamination than either of the alternative methods investigated in this paper. The ensemble configurations show the best overall performance, followed by the single-classifier configuration. Of the alternative 
methods, shape asymmetry outperforms WND-CHARM. All the methods tested definitively outperform random guessing. 

The completeness and contamination of the two CNN ensemble configurations are notably improved over the single CNN for the more challenging half of tidal galaxies, leveraging residual independence between classifiers to increase performance. For the clearest half of tidal galaxies, the network ensembles show relatively little improvement. This could be a consequence of there being relatively little disagreement between ensemble classifiers for the most obvious examples, reducing the impact of inter-classifier voting. 

We deem the shape asymmetry method to be moderately effective in identifying galaxies with faint tidal structure. For example, Figure \ref{allrocs} shows that it can achieve a completeness of 58\% with a contamination of 20\%.  As shown in Fig \ref{pawlik_probability}, there is generally a decent separation between tidal and non-tidal galaxies in $A_s$ versus $A$ parameter space. Galaxies with high values of both $A_s$ and $A$ are likely to be tidal, and for extreme values of these parameters the tidal prediction can be made confidently. This leads to the sharp gradient (corresponding to a rapid rise in completeness while preserving a low contamination) observed on Fig \ref{allrocs} for contamination $< 0.2$. However, the shape asymmetry method performs less well for galaxies with moderate $A_s$ and $A$, causing the gradient to subsequently flatten as less confident predictions are included. Extending shape asymmetry to use logistic regression rather than cuts provides only a small improvement.  

We also find that WND-CHARM is the least effective method investigated for identifying galaxies with faint tidal features in CFHTLS-Wide sample. We speculate that its poor performance may be a consequence of the macroscopic feature extraction step employed in the algorithm. The ratio of image information content corresponding to faint tidal features may be sufficiently low that WND-CHARM struggles to identify a genuinely predictive feature set amongst the `noise' of the general morphology. With an ability to investigate up to 4027 image features for correlations with labels, WND-CHARM could be overfitting to image features that do not relate to tidal features in the test data. However, WND-CHARM does show impressive performance on low confidence predictions, even exceeding that of shape asymmetry for contamination $\lesssim 0.5$. Ironically, this suggests that WND-CHARM is able to detect indications of tidal features on the more challenging images while it struggles to confidently detect such features on the clearest examples.

\section{Discussion}
\label{discussion}

\subsection{Heatmaps}
\label{heatmap}

A common criticism of CNNs, and deep learning in general, is that they are $`$black box' algorithms which are difficult to interpret.  While the resultant classification is readily apparent, the way in which it was arrived at is usually less so. 
There is no clear link from the properties of the galaxy features to the prediction made.

In order to establish if our method is really identifying faint tidal features in the way we intend, we use prediction heatmaps \citep{Zeiler2014}. Having established that each ensemble offers comparable performance, we arbitrarily investigate Configuration A (similar individual classifiers).

For a single image, we inject a synthetic low surface brightness tidal structure into a small area. First, we create a 5x5 grid of pixel values from a Gaussian distribution with background variance and a mean 3$\sigma$ above the background which represents the synthetic structure. Second, we take the original image and replace a random 5x5 pixel area with our new structure. 

Each time we add the structure, we reclassify the new image (original plus synthetic structure) with an ensemble classifier and record the change in tidal prediction from the original image. By plotting the tidal predictions as a heatmap where each pixel is the tidal prediction given a 5x5 synthetic structure at the location, we can identify in which image regions the ensemble sensitive to small changes. We assume that adding a tiny synthetic structure to a region that the network prediction is highly sensitive (one might say, `suspects' as being tidal) causes a much greater increase in the tidal prediction for the whole image than adding such structure to an otherwise non-tidal region.

\begin{figure}
  \includegraphics[width=\columnwidth]{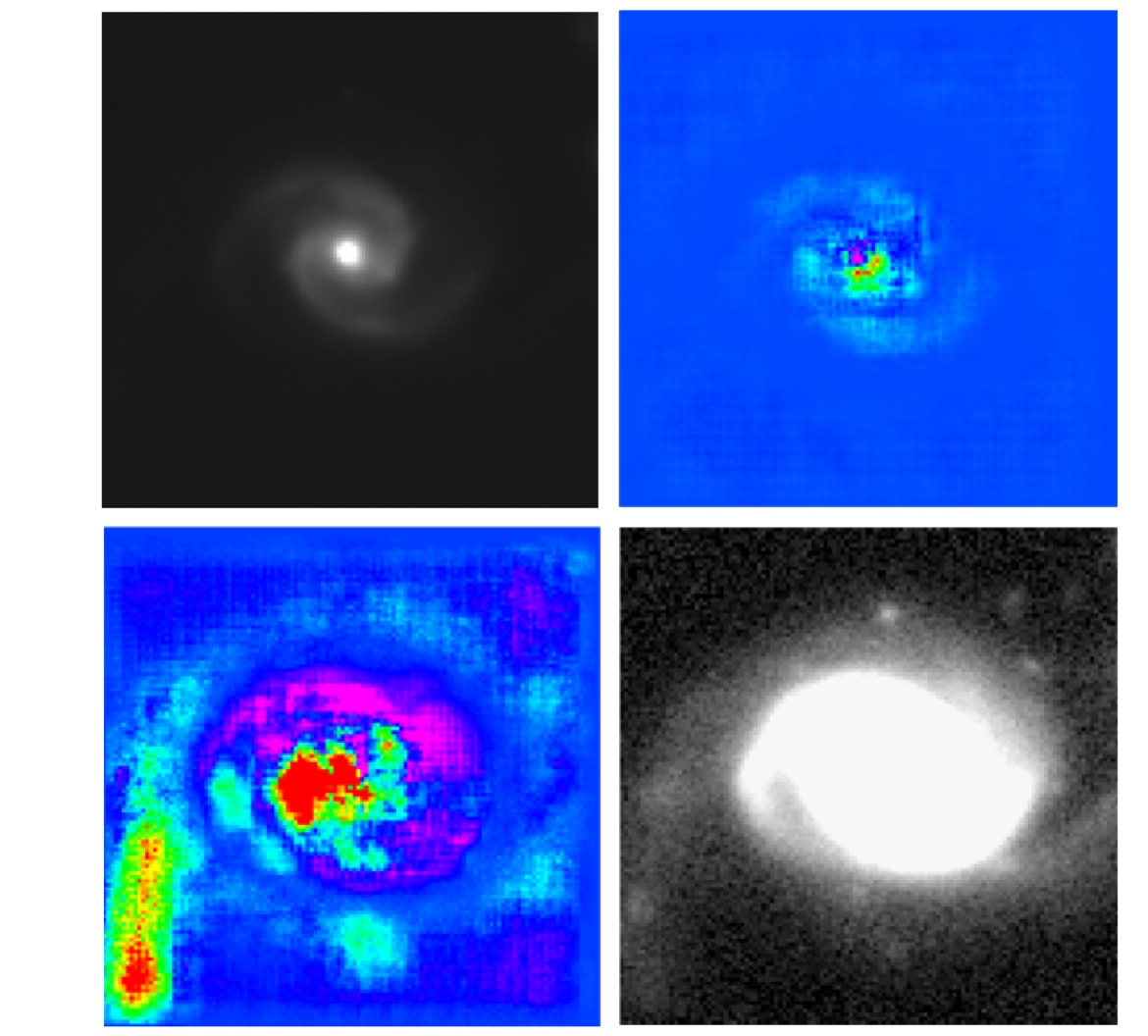}
  \caption{Top left: a cleaned galaxy image without rescaling. Top right: the heatmap from epoch 5. Bottom left: the heatmap from epoch 125. Bottom right: cleaned image with logarithmic rescaling. Magenta denotes non-tidal. Blue denotes neutral. Green through yellow through red denotes increasing tidal confidence. Note that the synthetic tidal structure is only added temporarily to alter the network predictions, and is not shown in any of the images above.}
  \label{heatmaps}
\end{figure}

Figure \ref{heatmaps} shows one example. The input image is shown at the top left. After a brief (5 epoch) training period, the heatmap is approximately a pixel-count-weighted distribution. After training is complete (epoch 125), the heatmap shows the network to have identified a linear feature at the bottom left corner of the image. Redisplaying the original image on a logarithmic scale, we verify that there is indeed a low surface brightness linear feature present at that location. This feature is detected and localised by the network despite being sufficiently faint to be invisible to the eye on the unscaled input image.  

Our prediction heatmap demonstrates that the CNNs are identifying which image pixels are associated with low surface brightness tidal features. If the pixel associations are sufficiently reliable, this offers the potential for automatic measurement of the shapes of tidal features as well.

\subsection{Training Data}
\label{training_data}

The sophistication of the CNNs used in this paper is limited by the size of the training data. The expert labels from A13 contain 305 tidal galaxies spread over six non-exclusive morphological classes of tidal feature. This places a fundamental limit on how much a convolutional network can generalise and learn to recognise such features. Pre-processing, shallow network design, augmentation and dropout are all necessary to achieve our classification performance.  

Larger training sets would provide constraining information to support CNNs with more free parameters. This in turn would allow for more complex predictions about the input images. In principle, a CNN could directly localise tidal features with bounding boxes (\citealt{Huang2016}), provide predictions for many different classes of tidal features (\citealt{Simonyan2014}), and estimate tidal parameters like the length of a tidal tail (\citealt{Toshev2014}). 
Our heatmap experiment (Section \ref{heatmap}) provides compelling evidence for the plausibility of these applications if a sufficiently large training sample can be realised. We discuss three possibilities for this below. 

Visually identifying large samples of galaxies with faint tidal structure is a daunting task given the relative rarity of such features at the typical surface brightness levels of current wide-field datasets.  Most studies agree that to a surface brightness of $\mu \sim 26.5-28$ mag arcsec$^{-2}$, roughly 10-20\% of galaxies show evidence for faint tidal features (e.g. A13, \citealt{Kado-Fong2018, Hood2018, Morales2018}. In order to create a training sample of even $\sim 10,000$ tidal systems, more than 100,000 galaxies would need to be visually inspected. Crowd-sourcing efforts like Galaxy Zoo \citep{Lintott2008, Willett2013} could  be an effective way to accomplish this but we note that tidal features from minor mergers and accretions are often rather subtle in appearance and visual identifications typically require some degree of interactive manipulation of pixel scaling and contrast. While citizen science may still prove an effective way forward, the accuracy of resulting tidal labels would need to be carefully verified, perhaps by checking against a smaller expert catalogue. 
Citizen science would also help mitigate the risk of a single expert producing classifications which systematically deviate from other experts.

Alternatively, or in conjunction, one could use synthetic training data from simulations. Individual tidal features can be simulated in exquisite detail (e.g. \citealt{Johnston2008, Hendel2015} and large-scale hydrodynamical simulations of galaxy formation now have the resolution to resolve these features in populations of several thousand galaxies (e.g. \citealt{Pop2018}).  With simulated data, mock observations could  be made at many viewing angles and surface brightness thresholds in order to provide an arbitrarily large training sample. However, while simulations provide perfect information on tidal labels, they are unlikely to fully capture the  development and evolution of real tidal features, impairing the ability of the classifier to detect such features.   

Finally, transfer learning provides an indirect method to include training data. First, a convolutional network is trained to solve a related problem on an independent training set. The convolutional layers of the network become able to extract features relevant to that related problem. Second, those feature-extracting layers are used to construct a new convolutional network aimed at solving the target problem.  The filters learned by those feature-extracting layers may be useful to re-apply. For example, learned filters that detect shapes and orientation on the related problem may be helpful for the target problem (see \citealt{NIPS2014_5347}). The features learned by CNNs trained on general galaxy morphology problems with far larger samples (\citealt{Willett2013,Dieleman2015,Huertas-Company2015a}) could be particularly relevant for detecting faint tidal features. A recent application of this is provided by \cite{Ackermann2018} who use transfer learning in conjunction with CNNs to automatically identify images of galaxy mergers.

\subsection{Application to New Data}

We ultimately aim to apply this method to detect tidal features in a large galaxy sample not previously classified. 
It is therefore important to ensure that that this method scales.

Each ensemble classifier makes tidal predictions on the order of 100 galaxies per second on a standard 2.4Ghz CPU, or approximately eight million galaxies per CPU-day. 
This means that classifying forthcoming samples from LSST and Euclid, which will be several orders-of-magnitude larger than the A13 sample, is computationally feasible. 

A13 manually removed images contaminated by stars, which would not be feasible for a large sample. 
However, automatic identification of contaminating stars is straightforward \citep{Soumagnac2015, Kennamer2018, Cabayol2018, Sevilla-Noarbe2018}. 
Current methods reach an AUC score exceeding 0.99 on comparable CFHT imaging \citep{Kim2017}.
For LSST-scale samples, we would use such methods to automatically remove contaminating stars prior to application of our convolutional neural network.

We removed as uninformative 8\% of images (136 of 1757) with expert labels of exactly 50\% confidence in tidal features. 
The performance metrics reported apply only to this slightly cleaner sample. 
Assuming classifiers guess randomly for such uncertain galaxies, and the true labels are equally random, the AUC scores of all the methods discussed would be slightly lower.
This does not affect our demonstration of the relative strength of convolutional neural networks at detecting tidal features.

\subsection{Potential Bias}
\label{bias}

Scalability is only meaningful if we understand the biases involved in the classifications. 
There are two important sources of bias introduced by the classifier that need to be considered. 

In the first case, the classifier may perform particularly poorly at recognising some classes of tidal features (e.g. streams or shells).
It is crucial to understand these biases so that they may be distinguished from genuine trends in the galaxy population. 
One way to approach this would be to construct a $`$calibration' catalogue where the true tidal feature labels are known. 
This could be achieved through using multi-expert visual classifications, or even synthetic data. 
Given a calibration catalogue, one can measure how classifier performance varies for each tidal feature class.
We measure the performance of our classifier by tidal debris class in Section \ref{singleresults}. 
Should some classes be poorly recognised, one could either apply an appropriate correction or search for additional examples of that tidal feature class to improve performance.

On the other hand, within any given dataset, bias may be triggered by the image context. 
Experts understand that they should not consider bright foreground or background objects, diffraction spikes or any other $`$artefacts'  when making a classification. 
CNNs have no such expertise unless inferred from the training data. 
Further domain-specific augmentations could help the classifier avoid confusion from these context biases.
Adding synthetic observational effects would provide training examples to teach the classifier to ignore such effects and better handle, for example, classifications of galaxies in crowded images.  

\section{Conclusion}
\label{conclusion}

We have examined the performance of CNNs with dropout and augmentation to identify galaxies in the CFHTLS-Wide Survey that have faint tidal features in their outer regions. Learning the ideal features to extract from the pixel data and gradually increasing the pixel scale of feature maps make CNNs effective at classifying features in complex images. We have shown that appropriate preprocessing and augmentation combined with a relatively shallow network architecture is key to avoiding overfitting of the data.  Randomised five-fold cross-validation verifies that our results are independent of which images are selected for training and which for testing. Training and testing five uniquely-instantiated CNNs in two different ensemble configurations confirms that our results are statistically reliable and do not result from a fortuitous instantiation of initial weights.  Through adding mock tidal features, we have shown that our method highlights image features that are found to be discriminatory without applying a parametric model. 

Comparing the performance of our classifiers against previously-published expert visual classifications,  we find that our method achieves high (76\%) completeness and low (20\%) contamination. It also performs considerably better than other automated methods recently applied in the literature, namely the shape asymmetry method, a non-parametric approach developed for identifying post-merger galaxies by \citealt{Pawlik2016}), and WND-CHRM, a generic machine learning approach previously applied to image classification in astronomy (\citealt{Shamir2012}). 

Our demonstration of the effectiveness of CNNs represents a significant step forward in developing a fully-automated method for faint tidal feature detection in galaxies. Indeed, most work in detecting and classifying tidal features in galaxies is still wholly or partially dependent on expert visual identification (e.g. \citealt{Kado-Fong2018, Hood2018, Morales2018}).  This  strategy is 
completely inadequate for the next generation of deep wide field surveys, such as LSST and Euclid, which will cover $\sim 15,000-20,000$ square degrees at unprecedented photometric depth \citep{Laureijs2011,Robertson2017}. While a limiting factor is the lack of currently-available training data, the use of either citizen science labels, simulation data or transfer learning are potential ways to address this. The development of a robust and efficient method to not only identify, but also characterise, faint tidal features around galaxies will enable the record of minor mergers and interactions to be mined in very large statistical samples. This will provide unique and previously inaccessible insight into the history of the galaxy population over cosmic time and facilitate the much-anticipated revolution that next generation facilities promise in terms of quantitative low surface brightness science. 

\section*{Acknowledgements}

We would like to thank the reviewer Lior Shamir for their helpful comments and suggestions.

MW acknowledges funding from the Science and Technology Funding Council (STFC) Grant Code ST/R505006/1. AMNF acknowledges support from STFC and the Alexander von Humboldt Foundation.

Based on observations obtained with MegaPrime/MegaCam, a joint project of CFHT and CEA/IRFU, at the Canada-France-Hawaii Telescope (CFHT) which is operated by the National Research Council (NRC) of Canada, the Institut National des Science de l'Univers of the Centre National de la Recherche Scientifique (CNRS) of France, and the University of Hawaii. This work is based in part on data products produced at Terapix available at the Canadian Astronomy Data Centre as part of the Canada-France-Hawaii Telescope Legacy Survey, a collaborative project of NRC and CNRS. 

This research made use of the open-source Python scientific computing ecosystem, including SciPy \citep{scipy}, Matplotlib \citep{Hunter2007}, scikit-learn \citep{Pedregosa2011}, scikit-image \citep{VanderWalt2014} and Pandas \citep{McKinney2010}.

This research made use of Astropy, a community-developed core Python package for Astronomy \citep{TheAstropyCollaboration2013,TheAstropyCollaboration2018}.

This research made use of the deep learning Python package Keras \citep{chollet2015keras}, recently included within TensorFlow \citep{tensorflow2015-whitepaper}.

All code is publicly available on Github at \path{www.github.com/mwalmsley/tidal_features_classifier} \citep{Walmsley2018}.


\bibliographystyle{mnras}
\bibliography{bibliography}


\bsp	
\label{lastpage}
\end{document}